%% file: main.tex
\documentclass{template/ar-1col}
\input{preamble}

\begin{document}

% Page header
\markboth{Drischler, Holt, and Wellenhofer}{Chiral EFT and the High-Density Nuclear EOS}

% Title
\title{Chiral Effective Field Theory and the High-Density Nuclear Equation of State}

%Authors, affiliations address.
\author{C.~Drischler,$^{1,2,3}$ J.W.~Holt,$^4$ and C.~Wellenhofer,$^{5,6}$
\affil{$^1$Department of Physics, University of California, Berkeley, California 94720, USA}
\affil{$^2$Nuclear Science Division, Lawrence Berkeley National Laboratory, Berkeley, California 94720, USA}
\affil{$^3$Facility for Rare Isotope Beams, Michigan State University, Michigan 48824, USA; email: \url{drischler@frib.msu.edu}}
\affil{$^4$Cyclotron Institute and Department of Physics and Astronomy, Texas A\&M University, College Station, Texas 77843, USA;
email: \url{holt@physics.tamu.edu}}
\affil{$^5$Institut f{\"u}r Kernphysik, Technische Universit{\"a}t Darmstadt,
64289 Darmstadt, Germany; email: \url{wellenhofer@theorie.ikp.physik.tu-darmstadt.de}}
\affil{$^6$ExtreMe Matter Institute EMMI, GSI Helmholtzzentrum f{\"u}r
Schwerionenforschung GmbH, 64291 Darmstadt, Germany}
}

%Abstract
\begin{abstract}
%Abstract text, approximately 150 words. [currently 145 words]
%Recent advances in neutron star observations have the potential to constrain the properties of strongly interacting %matter at extreme densities and temperatures that are otherwise difficult to access through direct experimental %investigation.

Born in the aftermath of core collapse supernovae, neutron stars contain matter under extraordinary conditions of density and temperature that are difficult to reproduce in the laboratory. In recent years, neutron star observations have begun to yield novel insights into the nature of strongly interacting matter in the high-density regime where current theoretical models are challenged. At the same time, chiral effective field theory has developed into a powerful framework to study nuclear matter properties with quantified uncertainties in the moderate-density regime for modeling neutron stars. In this article, we review recent developments in chiral effective field theory and focus on many-body perturbation theory as a computationally efficient tool for calculating the properties of hot and dense nuclear matter. We also demonstrate how effective field theory enables statistically meaningful comparisons between nuclear theory predictions, nuclear experiments, and observational constraints on the nuclear equation of state.

\end{abstract}

%Keywords, etc.

% Metadata Information
\jname{Annu. Rev. Nucl. Part. Sci.} 
\jvol{71} 
\jyear{2021} 
\doi{10.1146/annurev-nucl-102419-041903}
\firstpagenote{Annu. Rev. Nucl. Part. Sci. in press.}
%Draft compiled on \today.\newline Do not share or distribute.}

\maketitle

%Table of Contents, figures, ...
\tableofcontents
%\listoffigures  % not required by journal, remove prior to submission

%main text
\input{sections/01_introduction.tex}
\input{sections/02_mb_problem}
\input{sections/03_nuclear_matter.tex}
\input{sections/04_applications.tex}
\input{sections/05_summary_outlook.tex}
\section*{DISCLOSURE STATEMENT}

The authors are not aware of any affiliations, memberships, funding, or
financial holdings that might be perceived as affecting the objectivity of this
review.

% Acknowledgements
\section*{ACKNOWLEDGMENTS}

We are very grateful to our collaborators and colleagues for fruitful discussions over the years that contributed in one way or another to the completion of this review article. We specifically thank Evgeny Epelbaum for providing us with the diagrams in Figure~\ref{fig:EFT_table} and Yeunhwan Lim for sharing the results of Figures~\ref{fig:R14_S2} and~\ref{fig:press_ns}. C.D. acknowledges support by the Alexander von Humboldt Foundation through a Feodor-Lynen Fellowship and thanks the N3AS and BUQEYE Collaborations for creating warm research environments. This material is based upon work supported by the U.S. Department of Energy, Office of Science, Office of Nuclear Physics, under the FRIB Theory Alliance award DE-SC0013617. The work of J.W.H.\ was supported by the National Science Foundation under Grant No.\ PHY1652199 and by the U.S.\ Department of Energy National Nuclear Security Administration under Grant No.\ DE-NA0003841. C.W. has been supported by the  Deutsche Forschungsgemeinschaft (DFG, German Research Foundation) -- \mbox{Project-ID} 279384907 -- SFB 1245.

% References
%
% Margin notes within bibliography
\bibliographystyle{template/ar-style5}
\bibliography{bib/eft.bib}

\end{document}

%% file: preamble.tex
\usepackage{url}
\usepackage[numbers]{natbib}
\usepackage{graphicx}
\usepackage{amsfonts,amsmath,amssymb,bm}
\usepackage{xparse,xspace}
\usepackage{physics}
\usepackage[pdfpagelabels, pdfencoding=auto, psdextra]{hyperref}
\usepackage{subcaption}
\usepackage{isotope}
\usepackage{braket}
\usepackage{ulem}
\usepackage{mathrsfs}
\usepackage{mathtools}
\usepackage{feynmp-auto}
\usepackage{color}

\DeclareFontFamily{U}{FdSymbolA}{}
\DeclareFontShape{U}{FdSymbolA}{m}{n}{<-> s * FdSymbolA-Book}{}
\DeclareSymbolFont{fdmarkers}{U}{FdSymbolA}{m}{n}
\DeclareMathSymbol{\smallblacktriangleright}{\mathrel}{fdmarkers}{90}
\DeclareMathSymbol{\smallblacktriangleup}{\mathrel}{fdmarkers}{91}
\DeclareMathSymbol{\smallblacktriangleleff}{\mathrel}{fdmarkers}{92}
\DeclareMathSymbol{\smallblacktriangledown}{\mathrel}{fdmarkers}{93}
\DeclareMathSymbol{\smallblackcircle}{\mathrel}{fdmarkers}{114}
\DeclareMathSymbol{\smallblacksquare}{\mathrel}{fdmarkers}{128}
\DeclareMathSymbol{\smallblackdiamond}{\mathrel}{fdmarkers}{132}
\DeclareMathSymbol{\smallblackstar}{\mathrel}{fdmarkers}{151}

\definecolor{tab20:0}{rgb}{0.12156862745098039,0.4666666666666667,0.7058823529411765}
\definecolor{tab20:1}{rgb}{0.6823529411764706,0.7803921568627451,0.9098039215686274}
\definecolor{tab20:2}{rgb}{1.0,0.4980392156862745,0.054901960784313725}
\definecolor{tab20:3}{rgb}{1.0,0.7333333333333333,0.47058823529411764}
\definecolor{tab20:4}{rgb}{0.17254901960784313,0.6274509803921569,0.17254901960784313}
\definecolor{tab20:5}{rgb}{0.596078431372549,0.8745098039215686,0.5411764705882353}
\definecolor{tab20:6}{rgb}{0.8392156862745098,0.15294117647058825,0.1568627450980392}
\definecolor{tab20:7}{rgb}{1.0,0.596078431372549,0.5882352941176471}
\definecolor{tab20:8}{rgb}{0.5803921568627451,0.403921568627451,0.7411764705882353}
\definecolor{tab20:9}{rgb}{0.7725490196078432,0.6901960784313725,0.8352941176470589}
\definecolor{tab20:10}{rgb}{0.5490196078431373,0.33725490196078434,0.29411764705882354}
\definecolor{tab20:11}{rgb}{0.7686274509803922,0.611764705882353,0.5803921568627451}
\definecolor{tab20:12}{rgb}{0.8901960784313725,0.4666666666666667,0.7607843137254902}
\definecolor{tab20:13}{rgb}{0.9686274509803922,0.7137254901960784,0.8235294117647058}
\definecolor{tab20:14}{rgb}{0.4980392156862745,0.4980392156862745,0.4980392156862745}
\definecolor{tab20:15}{rgb}{0.7803921568627451,0.7803921568627451,0.7803921568627451}
\definecolor{tab20:16}{rgb}{0.7372549019607844,0.7411764705882353,0.13333333333333333}
\definecolor{tab20:17}{rgb}{0.8588235294117647,0.8588235294117647,0.5529411764705883}
\definecolor{tab20:18}{rgb}{0.09019607843137255,0.7450980392156863,0.8117647058823529}
\definecolor{tab20:19}{rgb}{0.6196078431372549,0.8549019607843137,0.8980392156862745}

\hypersetup{%
    pdfsubject=Paper,
    pdfkeywords={nuclear physics} {Bayesian} {chiral EFT} {nuclear matter},
    unicode = true,
    breaklinks = true,
    colorlinks = true,
    linkcolor = blue,
    menucolor = blue,
    citecolor = blue,
    urlcolor = blue
}

\allowdisplaybreaks  % did not come with the journal template
\graphicspath{{./figs/}}

\input{macros}

%% file: macros.tex
\newcommand{\etal}{\textit{et~al.}\xspace}
\newcommand{\eg}{\textit{e.g.}\xspace}
\newcommand{\ie}{\textit{i.e.}\xspace}

\newcommand{\Msun}{\,\text{M}_{\odot}\xspace}

\newcommand{\km}{\, \text{km}}

\newcommand{\fmi}{\, \text{fm}^{-1}}
\newcommand{\fmiq}{\, \text{fm}^{-3}}

\newcommand{\MeV}{\, \text{MeV}}
\newcommand{\GeV}{\, \text{GeV}}

\newcommand{\nsat}{\, n_0}

\newcommand{\NNLO}{\ensuremath{{\rm N}{}^2{\rm LO}}\xspace}
\newcommand{\NNNLO}{\ensuremath{{\rm N}{}^3{\rm LO}}\xspace}

\newcommand{\NkLO}[1]{\ensuremath{\mathrm{N}^{#1}\mathrm{LO}}\xspace}

\newcommand{\NNLOsat}{\ensuremath{\mathrm{NNLO}_{\mathrm{sat}}}\xspace}

\newcommand{\ChEFT}{ChEFT\xspace}
\newcommand{\kF}{\ensuremath{k_{\scriptscriptstyle\textrm{F}}}}

\newcommand{\e}{e} %% CD: can this trivial command be removed?
\newcommand{\bs}[1]{\boldsymbol{#1}}

\newcommand{\genobs}{y}
\newcommand{\inputdimvec}{}
\newcommand{\kinparvec}{\inputdimvec{n}}
\newcommand{\genobsref}{{y_{\mathrm{ref}}}}
\DeclareMathOperator{\pr}{pr}

\newcommand{\GPB}[1]{\mbox{GP--B~#1}}

%% file: sections/01_introduction.tex
\section{Introduction} \label{sec:intro}

Neutron stars are one of Nature's most intriguing %stellar 
astronomical objects and provide a unique laboratory for studying strongly interacting, neutron-rich matter under extreme conditions. With masses of about $1-2$ times that of the Sun and radii of only approximately $10\km$, neutron stars contain the densest form of matter in the observable Universe and lie just at the threshold for collapse to a black hole. Much has already been learned about neutron stars through mass and radius measurements, pulsar timing, x-ray observations, and gravitational-wave measurements of binary mergers in the new era of multimessenger astronomy (see, \eg, Refs.~\cite{OzelARNAA,Watts16,baiotti19} for reviews). But many interesting questions remain to be answered, especially regarding the nature of ultra-compressed matter located in the inner cores of heavy neutron stars where a variety of exotic new states of matter have been theorized to exist.
%
%\begin{marginnote}
%The mass of the Sun is $\Msun \approx 2 \times 10^{30} \kg$.
%\end{marginnote}

While neutron stars are bound together by gravity acting over macroscopic length scales, strong short-ranged nuclear interactions provide the essential pressure support to counteract gravitational collapse. The central densities in the heaviest neutron stars may reach up to $5-10\nsat$, where $n_0\approx 0.16\fmiq$ is the nucleon number density typical of heavy atomic nuclei
(the associated mass density is 
$\rho_0 \approx 2.7 \times 10^{14} \, \text{g\,cm}^{-3}$).
Although the strong interaction is in principle described by quantum chromodynamics (QCD) over all relevant energy scales, 
at present no systematic computational method is available to calculate 
the properties of the high-density matter in the inner cores of heavy neutron stars.
\begin{marginnote}
	\entry{QCD}{quantum chromodynamics}
	\entry{\ChEFT}{chiral effective field theory}
\end{marginnote}
With chiral effective field theory (\ChEFT), however, a powerful tool has emerged to carry out microscopic calculations of nuclear matter properties at densities up to around $2n_0$. 
Instead of QCD's quarks and gluons, \ChEFT is formulated in terms of nucleons and pions (and delta isobars), which are the effective strong interaction degrees of freedom present throughout most of the neutron star interior. 
In its range of validity, \ChEFT
provides a systematic expansion for two- and multi-nucleon interactions consistent with the symmetries of low-energy QCD.
%, in particular the spontaneously broken chiral symmetry. 
%In the prevalent way of organizing this expansion, degrees of freedom at and above the chiral symmetry breaking scale $\Lambda_\chi \sim 1 \GeV$ are explicitly integrated out in the construction of the nuclear force, which can then be implemented in any number of many-body frameworks. 
The unresolved short-distance physics is parameterized in terms of contact interactions whose low-energy couplings are fitted to experimental data.
%Applying chiral nuclear interactions in a given many-body framework provides improvable predictions for nuclear-matter properties.
An essential advantage over phenomenological approaches is that theoretical uncertainties can be quantified by analyzing the order-by-order convergence of the \ChEFT expansion. 
In the last few years, the combination of systematic nuclear matter predictions from \ChEFT, uncertainty quantification, and neutron star observations 
has developed into a new avenue for constraining the high-density regime of the nuclear equation of state (EOS).   
\begin{marginnote}
%	\entry{LEC}{low-energy constant}
	%\entry{NN}{nucleon-nucleon}
	%\entry{3N}{three-nucleon}
	%\entry{4N}{four-nucleon}
	\entry{EOS}{equation of state}
	\entry{MBPT}{many-body perturbation theory}
\end{marginnote}

%\begin{marginnote}
%	\entry{SNM}{symmetric nuclear matter}
%	\entry{PNM}{pure neutron matter}
%\end{marginnote}

In this review our aim is to describe recent advances in microscopic \ChEFT calculations of the nuclear EOS and their application to neutron stars. [For recent reviews of nuclear structure calculations with \ChEFT see, \eg, Refs.~\cite{Hergert:2020bxy,Stroberg:2019mxo,Lynn:2019rdt,Hebeler:2015hla}.] We highlight many-body perturbation theory (MBPT) as an efficient framework for nuclear matter calculations at zero and finite temperature based on chiral two- and multi-nucleon interactions. We also discuss Bayesian methods for quantifying and propagating statistically meaningful theoretical uncertainties. Together with nuclear experiments, astrophysical simulations, and neutron star observations, next-generation \ChEFT calculations will be crucial to infer the nature of the extreme matter hidden deep beneath the surface of neutron stars.
The review is organized as follows. In Section~\ref{sec:mb_problem} we focus on recent progress in deriving nuclear forces from \ChEFT and renormalization group (RG) methods to improve the many-body convergence in nuclear matter calculations.
%Further, we provide a concise introduction to MBPT at zero and finite temperature, and compare its efficacy with that of complementary nonperturbative many-body methods, in particular regarding uncertainty quantification issues.
We then dedicate Section~\ref{sec:nuclear_matter} to recent high-order MBPT calculations of the moderate-density nuclear EOS at zero temperature and advances in the Bayesian quantification of EFT truncation errors. We also discuss finite-temperature calculations and nuclear thermodynamics. In Section~\ref{sec:applications} we review the present status of the high-density nuclear EOS constrained by nuclear theory, experiment, and observation in the era of multimessenger astronomy, emphasizing the importance of \ChEFT. Section~\ref{sec:summary_outlook} ends the review with our summary and perspectives on future advances in nuclear matter calculations and their applications to astrophysics. 
\begin{marginnote}
%	\entry{CC}{Coupled-Cluster}
%	\entry{QMC}{Quantum Monte Carlo}
%	\entry{SCGF}{Self-Consistent Green's Functions}
	\entry{RG}{renormalization group}
\end{marginnote}

%\begin{afterthoughts}[Afterthoughts: remove prior to submission]
%	\begin{enumerate}
%		\item raise questions about chiral EFT, high-density EOS, etc. and mention that these will be addressed in the review.
%		\item be more specific on multimessenger astronomy; could be done elegantly by referring to Lattimer's review in the same issue
%		\item too much text?
%		\item mention normal ordering; more emphasis on 3N forces in general?
%		\item add here some good questions that sort-of guide today's research: Perhaps 1) Where does chiral EFT breakdown? What density does that breakdown scale correspond to? Do we need to include the delta at saturation density and beyond? How accurately do we need to constrain the EOS microscopically at low density in order to constrain (say) astro-observables at the 10 or 20\% level? To which extent do observational and experimental constraints agree with microscopic calculations?
%\end{enumerate}
%\end{afterthoughts}

%% file: sections/02_mb_problem.tex
%\section{Nuclear many-body problem: from microscopic interactions to the EOS} \label{sec:mb_problem}
\section{From microscopic interactions to the nuclear equation of state} \label{sec:mb_problem}

%In this section we aim to provide an overview of modern nuclear-matter calculations from the perspective of MBPT. We start with briefly reviewing (delta-less) \ChEFT in general as well as recent advances in the construction of chiral interactions. These are the microscopic input in the many-body calculations discussed in this review. The many-body convergence can often be improved by applying RG techniques to generate (perturbative) low-momentum interactions, for which the nuclear EOS and related observables can be efficiently calculated using MBPT. We discuss zero-temperature MBPT and its generalization to finite temperatures, complementary approaches, and the implementation of 3N interactions in many-body frameworks. 

In this section, we briefly review delta-less \ChEFT and the construction of chiral nuclear interactions as microscopic input for 
many-body calculations.
%the many-body frameworks. 
Applying RG methods allows one to systematically generate (perturbative) low-momentum interactions, for which the nuclear EOS and related observables can be efficiently calculated using MBPT.
We discuss both zero- and finite-temperature MBPT, complementary many-body approaches, and the implementation of 3N interactions in nuclear matter calculations.

%%%%%%%%%%%%%%%%%%%%%%%%%%%%%%%%%%%%%%%%%%%%%%%%%%%%%%
\subsection{Chiral effective field theory for nuclear forces} %\label{sec:chial_eft}
\label{sec21}

The interactions among nucleons arise as an effective low-energy phenomenon of QCD, the theory of the strong interaction.
At the momentum scales relevant for nuclear physics, $p \sim m_\pi$, QCD is strongly coupled and features nonperturbative effects such as spontaneous chiral symmetry breaking and the confinement of quarks and gluons into hadrons. Direct applications of QCD to hadronic physics at finite density, where lattice QCD faces a formidable sign problem, are therefore extremely challenging and not feasible at present and in the near future.
However, one can construct a systematic description of nuclear physics in terms of the effective degrees of freedom at low energies: nucleons and pions (and delta isobars).
This effective description is given by \ChEFT~\cite{Epel09RMP, Mach11PR,
	Hammer:2019poc, Hammer:2012id,Epelbaum:2019kcf}.

The starting point of \ChEFT is to write down the most general Lagrangian consistent with the symmetries of low-energy QCD, 
in particular, the spontaneously broken chiral symmetry, for which pions are the (pseudo) Nambu-Goldstone bosons.
This naturally sets a limit for the applicability of \ChEFT, \ie, 
the breakdown scale $\Lambda_b$ will be of order of the chiral symmetry breaking scale $\Lambda_\chi \sim 1 \GeV$. [A Bayesian analysis of free-space NN scattering with several (not-too-soft) chiral NN potentials in Weinberg power counting estimated $\Lambda_b \approx 600 \MeV$~\cite{Furnstahl:2015rha}.]
A truncation scheme, known as power counting, is then needed to organize the 
infinite number of operators in the effective Lagrangian in a systematic expansion.
%for deriving nuclear forces (and external currents).
This expansion is governed by the \textit{separation of scales} inherent in \ChEFT, \ie, 
%the power counting is according to 
%in powers of a typical momentum (or the pion mass) over the EFT breakdown scale, $Q=\max(q,m_\pi)/\Lambda_b$. 
the power counting is based on %according to 
powers of 
%the soft momentum scale $p$ (or the pion mass)
a typical momentum $p$ (or the pion mass)
over the \ChEFT breakdown scale, 
$Q=\max(p,m_\pi)/\Lambda_b$.
%
% \begin{marginnote} 
% A Bayesian analysis of free-space NN scattering with several (not-too-soft) chiral NN potentials in Weinberg power counting estimated $\Lambda_b \approx 600 \MeV$~\cite{Furnstahl:2015rha}.
% \end{marginnote}

In perturbative EFT, both power counting and ultraviolet renormalization are essentially unambiguous and straightforward.
The situation is different for applications of \ChEFT in nuclear physics, where the calculational framework must be 
able to account for nonperturbative effects
such as bound states (atomic nuclei) and large $S$-wave scattering lengths in NN scattering.
While there has been some controversy in the literature as to 
%the precise way 
how the \ChEFT expansion should be set up precisely (see, \eg, Refs.~\cite{Hammer:2019poc,Epelbaum:2019kcf} and references therein),
the prevalent and most successful power counting for \ChEFT (in particular regarding many-body applications) is the one first suggested by Weinberg~\cite{Weinberg:1990rz,Weinberg:1991um,Weinberg:1992yk}
\begin{marginnote}
	\entry{NN}{nucleon-nucleon}
	\entry{3N}{three-nucleon}	
	\entry{4N}{four-nucleon}
\end{marginnote}

%here brief general overview of ChEFT potential approach
Within Weinberg power counting, chiral nuclear interactions (and %external 
currents) are 
organized according to naive (\ie, perturbative) dimensional analysis. 
The nuclear potentials constructed at a given truncation order 
in the \ChEFT expansion are then used for computing observables.
Renormalization in this approach is approximative, and carried out by equipping the potentials with regulator functions that suppress contributions above a cutoff scale $\Lambda\lesssim \Lambda_b$, typically chosen in the range $450-600 \MeV$.
That is, the cutoff independence of the observables will be achieved only approximatively 
through $\Lambda$-dependent low-energy constants (LECs),
which are in practice optimized for a given $\Lambda$ to reproduce low-energy NN scattering data and few-nucleon observables; see, \eg, Refs.~\cite{Epelbaum:2019kcf,Reinert:2017usi}.
The residual cutoff dependence can then be attributed to higher-order terms in the 
%\ChEFT 
expansion, 
so results are %formally 
expected to become less cutoff dependent with increasing truncation order.
\begin{marginnote}
	\entry{LEC}{low-energy constant} 
% are in practice optimized for a given value of
% $\Lambda$ to reproduce low-energy NN scattering data and 
% few-nucleon observables, see, \eg, Refs.~\cite{Epelbaum:2019kcf,Reinert:2017usi}.}
\end{marginnote}

\begin{figure}[tb] 
	\includegraphics[width=\textwidth]{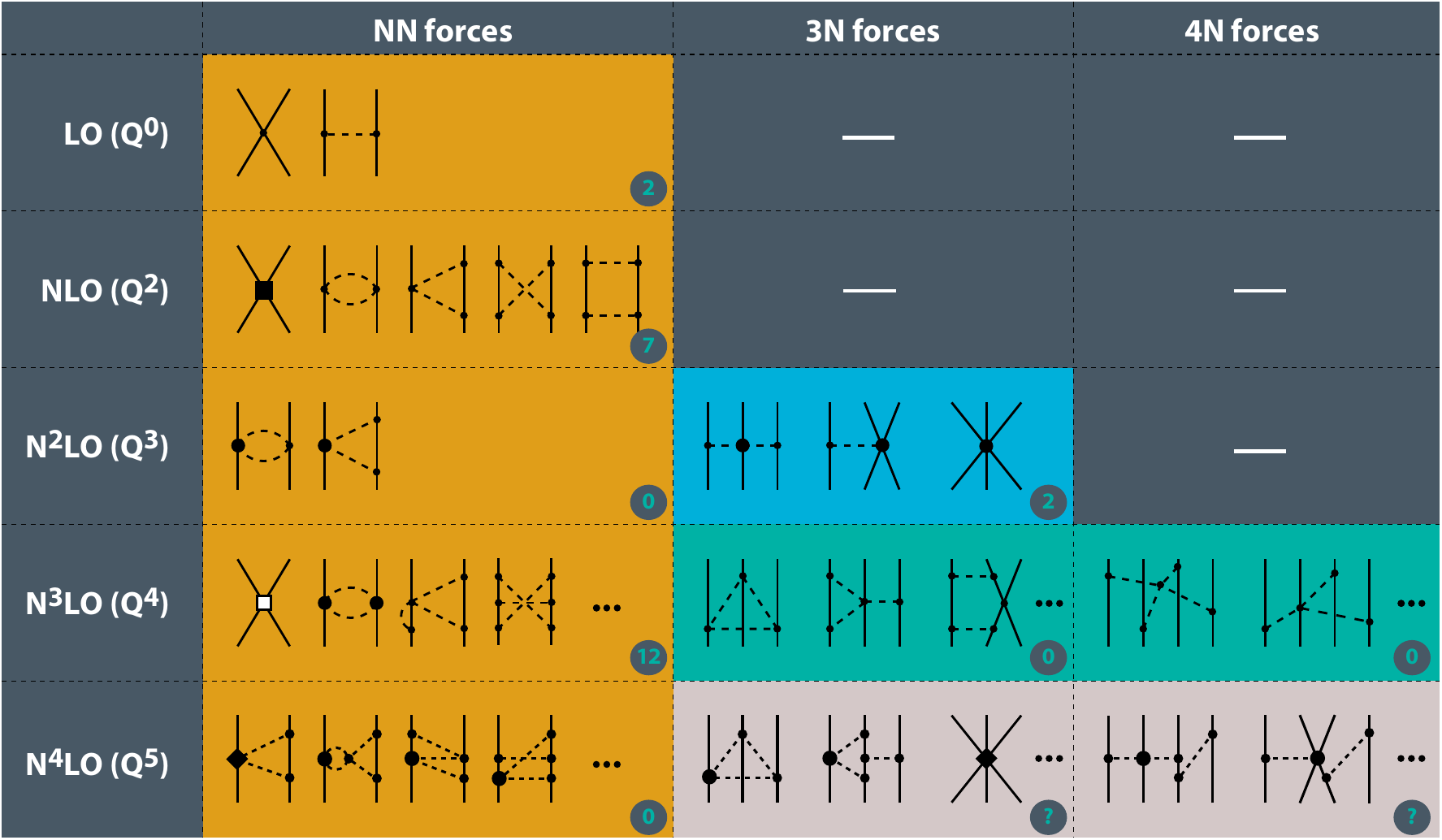}
		\caption[Chiral EFT table]{Hierarchy of chiral nuclear interactions up to fifth order (or \NkLO{4}) in the chiral expansion without delta isobars~\cite{Epelbaum:2019kcf}. Nucleons (pions) are depicted by solid (dashed) lines. The circled numbers give the number of short-range contact LECs. %See the main text for details. %The figure has been modified from Ref.~\addcite{}.
		}
	\label{fig:EFT_table}
\end{figure}
%%%%%%%%%%%%%%%%%%%%%%%%%%%%%%%%%%%%%%%%%%%%%%%%%%%%%%%%%%%%

%%here some details on LECs, NN 3N hierarchy etc
Figure~\ref{fig:EFT_table} depicts the hierarchy of nuclear interactions up to fifth order (or \NkLO{4}) in the chiral expansion 
without delta isobars. %[\ChEFT with explicit delta isobars is currently less developed than the delta-less version we focus on here. 
[For recent work on the currently less developed delta-full \ChEFT, see, \eg, Refs.~\cite{Jiang:2020the,Piarulli:2020mop}.] 
In this review, \NkLO{k} indicates (next-to)$^k$-leading order,
where $k$ is the number of orders beyond leading order (LO).
At each order the interactions are composed of short-range contact interactions as well as one- and multi-pion exchanges at long- and intermediate distances, respectively.
The LECs associated with pion exchanges 
have recently been determined with high precision 
through an analysis of pion-nucleon scattering
within the framework of Roy-Steiner equations~\cite{Hoferichter:2015tha}.
The short-range LECs corresponding to NN couplings are generally fixed by matching to NN scattering data.
Figure~\ref{fig:EFT_table} shows that \ChEFT naturally predicts the observed hierarchy of two- and multi-nucleon interactions, \ie, $V_\text{NN} > V_\text{3N} > V_\text{4N}$, etc.
The first nonvanishing 3N forces appear at \NNLO in three topologies; from left to right: the long-range two-pion exchange (involving the pion-nucleon LECs $c_{1}$, $c_{3}$, and $c_{4}$), intermediate-range one-pion exchange-contact ($\propto c_D$), and short-range 3N contact interaction ($\propto c_E$). At \NNNLO the 3N forces are significantly more involved and operator-rich, and also 4N interactions start to contribute. Apart from the two \NNLO 3N LECs, $c_D$ and $c_E$, chiral interactions up to \NNNLO are completely determined by the $\pi$N and NN system. While \NkLO{4} NN forces have already been worked out, partly even at \NkLO{5}, the derivation of \NkLO{4} 3N interactions has not been finished yet.
The 3N LECs $c_D$ and $c_E$ can be fit to (uncorrelated) few-body observables; for instance, the \isotope[3]{H} binding energy combined with, \eg, the charge radius of \isotope[4]{He}, the \isotope[3]{H} $\beta$-decay half-life, or the nucleon-deuteron scattering cross section. Also heavier nuclei and even saturation properties in infinite nuclear matter have been used to constrain 3N forces.
\begin{marginnote}
% 	\ChEFT with explicit delta isobars is currently less developed than the delta-less version we focus on here. For recent work on delta-full \ChEFT, see, \eg, Refs.~\cite{Jiang:2020the,Piarulli:2020mop}. %Strohmeier:2020dkb
		\entry{N\textsuperscript{k}LO}{(next-to)$^k$-leading order}
\end{marginnote}

%As noted, chiral nuclear potentials are constructed with a regulator function that suppresses momenta beyond a scale $\Lambda$, where typically $\Lambda$ is chosen in the range $450-600 \MeV$. The residual dependence of obervables on the functional form of the regulator and the value of $\Lambda$ is formally expected to be of higher order with respect to the truncation of the chiral expansion, but the actual form of regulator artifacts depends on the specific regularization method and the computational framework.

%The residual dependence of obervables on the functional form of the regulator and the value of $\Lambda$ is formally expected to be of higher order with respect to the truncation of the chiral expansion, but the actual form of regulator artifacts depends on the specific regularization method and computational framework.
Although the residual regulator and cutoff dependence of obervables at a given chiral order is expected to decrease at higher orders, actual calculations show significant influence of these so-called regulator artifacts on the \ChEFT convergence depending on the specific regularization scheme and computational framework. These issues have resulted in the development of a flurry of chiral potentials with nonlocal, local as well as semilocal regulators 
for a range of cutoff values; see, \eg, Table~I of Ref.~\cite{Hoppe:2017lok}.
Moreover, as discussed in Section~\ref{sec22}, RG methods allow one 
to modify a given set of two- and multi-nucleon potentials
such that observables are left invariant (up to RG truncations)
but the convergence of many-body calculations is optimized.
These RG transformations are most suitably formulated
at the operator (\ie, Hamiltonian) level. The nuclear Hamiltonian constructed at a given order in the \ChEFT expansion reads
$H = T_\text{kin} + V_\text{NN}(\Lambda,c_i) + V_\text{3N}(\Lambda,c_i) + V_\text{4N}(\Lambda,c_i) + \ldots\,,$
where $\Lambda$ stands for the (initial) cutoff or resolution scale, 
and $c_i$ for the set of LECs inferred from fits to experimental data.
The nuclear Hamiltonian is not an observable, and 
the basic idea of the RG is 
to exploit this feature to generate more perturbative Hamiltonians.

\subsection{Perturbative chiral nuclear interactions}
\label{sec22}

The strong short-range repulsion (``hard core'') and 
%strong 
tensor force found in nuclear potentials constructed at cutoff scales $\Lambda\gtrsim 500 \MeV$ raise questions regarding the applicability of perturbation theory for many-body calculations. In fact, nuclear many-body calculations were historically considered a nonperturbative problem (see also Section~\ref{sec23}). Both features give rise to strong couplings between high- and low-momentum states, \ie, large off-diagonal matrix elements, which enhance the intermediate-state summations 
%at high orders 
in perturbation theory. RG methods allow one 
to amend this feature while preserving 
nonperturbative few-body results.

%Nuclear potentials are not observable. There are infinitely many potentials that equally describe low-energy observables such as (nearly) bound states,  but differ in their momentum-dependence because short-range repulsion and tensor forces are resolution-dependent. For example, chiral interactions suppress these sources of nonperturbative behavior for low momentum cutoffs, $\Lambda \lesssim 500 \MeV$, in contrast to much harder phenomenological potentials like Argonne~$v_{18}$. The RG formalizes the resolution dependence of nuclear potentials and their evolution towards lower scales to systematically soften nuclear interactions.

The initial application~\cite{bogner01} of RG methods to study the scale dependence of nuclear forces was based on $T$-matrix equivalence, but in recent years the similarity renormalization group (SRG) has been the standard RG method for ``softening'' nuclear interactions. The SRG decouples high- and low-momentum states through continuous infinitesimal unitary transformations, $H_s = U_s H U_s^\dagger$, described by a differential flow equation in the evolution parameter~$s$. 
As the SRG flow progresses, the matrix-elements of the NN potential are driven toward a band-diagonal (or block-diagonal) form in momentum space, see Ref.~\cite{Bogner:2009bt} for illustrations.
While NN observables are by construction invariant under any RG evolution of the NN potential, 
$A$-body observables will remain so
only if one also consistently evolves the multi-nucleon part of the nuclear Hamiltonian.
The SRG allows one to implement this in principle exactly, in terms of so-called ``induced'' many-body forces. However, for practical applications, a truncation of the consistent evolution of multi-nucleon interactions is required, \eg, at the 3N level.
Neutron matter calculations with SRG-evolved chiral interactions truncated at the 3N level indicate that 
induced 4N forces are (in that case) negligible within uncertainties for a wide range of SRG resolution scales~\cite{PhysRevC.87.031302,1829502}.
\begin{marginnote}
	%\entry{RG}{Renormalization-group}
	\entry{SRG}{similarity renormalization group}
\end{marginnote}

There have been many developments recently in the application of \ChEFT and SRG technology for the construction of high-precision nuclear potentials.
Hebeler~\etal~\cite{Hebeler:2010xb} explored a set of low-momentum \NNNLO NN potentials combined with unevolved \NNLO 3N forces where the two 3N LECs were fit to reproduce few-body data (assuming that the 3N contact interactions %are able to
capture dominant contributions from induced 3N forces). 
For the softest of these potentials (with $\lambda = 1.8$ and $\Lambda_\text{3N} = 2.0 \fmi$), which was found to
predict nuclear saturation properties~\cite{Hebeler:2010xb,Drischler:2017wtt} and ground-state energies of light- to medium-mass nuclei in agreement with experiment~\cite{Simonis:2017dny}, 
Stroberg~\etal~\cite{Holt:2019gmc} have computed ground-state and separation energies of nearly 700 isotopes up to iron.
Moreover, H{\"u}ther~\etal~\cite{Huther:2019ont} have constructed a family of SRG-evolved NN and 3N potentials up to \NNNLO.  Also, Reinert~\etal~\cite{Reinert:2017usi} have developed the first chiral NN potentials up to \NkLO{4} with semilocal regulators in momentum space such that the long-range part of the pion exchanges remains invariant in contrast to nonlocal regulators. They showed that several \NNNLO contact terms present in previous generations of chiral NN potentials can be eliminated using unitary transformations, leading to considerably softer potentials (even without SRG evolution). 
The Weinberg eigenvalue analysis~\cite{PhysRev.131.440,Hoppe:2017lok} is a powerful tool for quantifying and monitoring the perturbativeness of nuclear forces at different resolution scales.
%For a 
Given an NN potential, 
the Weinberg eigenvalues $\eta_\nu(W)$ of the operator $G_0(W) V_\text{NN}$
determine the (rate of) convergence of the Born series for NN scattering. 
%in free space or a nuclear medium.
Here, $G_0(W)$ is the (free-space or in-medium) propagator as a function of the complex energy~$W$. 
The Born series converges if and only if 
all eigenvalues satisfy $|\eta_\nu(W)| < 1$.
Bound states of the potential (such as the deuteron) correspond to %energies 
%$W<0$ with $\eta_\nu(W) = 1$, 
$\eta_\nu(W) = 1$ at energies $W < 0$, so the free-space Born series diverges even for soft potentials. In nuclear matter at sufficiently high densities, however, Pauli blocking suppresses the (in-medium) eigenvalues associated with bound (or nearly-bound) states. For 
%low-momentum 
potentials constructed at $\Lambda \lesssim 550 \MeV$, other sources of nonperturbative behavior (such as the hard core) are suppressed as well, both in free-space and in-medium (see, \eg, 
%Section~2.4 in 
Ref.~\cite{Bogner:2009bt} for details).
%other eigenvalues inhibiting the convergence of the (free-space or in-medium) Born series are suppressed as well.
%This sets the stage for controlled MBPT calculations of the nuclear EOS.
This implies that a nonperturbative treatment of in-medium NN scattering in the particle-particle channel (see Section~\ref{sec23}) is not mandatory for these interactions.
Instead, order-by-order MBPT calculations can be used 
%as a systematic tool for studying
for systematically studying
the many-body convergence of (low-momentum) chiral nuclear interactions.
\subsection{Many-body perturbation theory at zero and finite temperature}
\label{sec23}

%This sections provides a brief review of MBPT, focusing in particular on its consistent generalization to finite temperatures. Compared to other methods, MBPT has the benefit that (at low orders) its computational cost is comparatively low. 
%%Further, the exact implementation of 3N forces in MBPT is straightforward, although at high orders ($N\gtrsim 4$) this increases the computational cost significantly.
%In the following, we restrict the discussion to NN interactions and deal with details regarding 3N interactions in Sec.~\ref{sec25}.

MBPT starts with partitioning the nuclear Hamiltonian $H$
%$H=T_\text{kin}+V$ 
into a 
reference one-body part $H_0=T_\text{kin}+U$ and a perturbation $H_\text{1}=V-U$, 
where $T_\text{kin}$ is the kinetic-energy operator and $U$ is an effective single-particle potential.
We consider here NN-only potentials %$V$ 
and discuss the implementation of 3N interactions in Section~\ref{sec25}.
The standard choice for $U$ is the Hartree-Fock potential\footnote{In Hartree-Fock MBPT, the $-U$ part of $H_\text{1}=V-U$ cancels all diagrams involving single-vertex loops~\cite{szabo,Drischler:2017wtt,1829502}.} given by
%\begin{align}
	$U^\text{(HF)}_{i} = \sum_{j} V^{ij,ij} f_{j}$,
%\end{align}
%in a shorthand notation,
with the antisymmetrized NN matrix elements $V^{ij,ab}=\braket{\vb{k}_i\vb{k}_j|(1-P_{12})V_\text{NN}|\vb{k}_a\vb{k}_b}$, the Pauli exchange operator $P_{12}$, the momentum integral $\sum_j=\int \dd[3]{k_j}/(2\pi)^3$, %and the distribution function of the reference state. %, \eg, 
and the zero-temperature distribution function $f_j=\theta(\kF-k_j)$.
For simplicity, 
%here and in the following 
we assume here a 
single-species system
%system with a single particle species 
and neglect spin-isospin degrees of freedom.
In zero-temperature MBPT, the %observable of interest, \eg, the 
ground-state energy density $\mathcal{E}$ %of the system
is obtained by expanding 
%in terms of 
$H_\text{int}$
about its reference value $\mathcal{E}_0$. %;
%see, \eg, Ref.~\cite{Fetter} for details.
Truncating the many-body expansion at a finite order $L$ then leads to the approximation
%\begin{align} \label{eq:MBPT_T0}
$\mathcal{E}(\kF)\simeq \mathcal{E}_0(\kF)+\sum_{l=1}^L \mathcal{E}_l(\kF)$,\footnote{The many-body expansion is in fact a divergent asymptotic series, 
but the divergent behavior is expected to appear only for 
high truncation orders $L\gtrsim 20$~\cite{Marino2019b}.}
%\end{align}
where the Fermi momentum $\kF$ is in one-to-one correspondence with the particle number density via
$n(\kF)=\sum_i f_i(\kF)$.
%
% \begin{marginnote}
% The MBPT series is in fact a divergent asymptotic series, 
% but the divergent behavior is expected to appear only for 
% high truncation orders $L\gtrsim 20$~\cite{Marino2019b}.
% %Thus, for all practical purposes MBPT (with low-momentum interactions) can be considered convergent.
% \end{marginnote}

%%%MBPT diagrammatics discussion
The first-order correction is determined by the expectation value of $U^\text{(HF)}_{i}$~\cite{szabo}.
At higher orders it is useful to represent the contributions diagrammatically, \eg, as Hugenholtz diagrams.
%We show the diagrams that appear at second and third order in Fig.~\ref{fig:manybodydiags}.
The diagram and expression for the second-order contribution $\mathcal{E}_2$ are given by
\begin{align}
	\mathcal{E}_{2}(\kF)=&
	\parbox[c][30pt]{40pt}{\input{diagrams/2nd_order}} = 
	-\frac{1}{4}\sum_{ijab}
	V^{ij,ab}V^{ab,ij} f_{ij}\bar f_{ab} \frac{1}{D_{ab,ij}}\,, \label{eq:E2kf}
\end{align} 
with the distribution functions $f_{ij}=f_if_j$ (holes) and $\bar f_{ab}=(1-f_a)(1-f_b)$ (particles), energy denominator $D_{ab,ij}=\varepsilon_a+\varepsilon_b-\varepsilon_i-\varepsilon_j$, and single-particle energies $\varepsilon_i=k_i^2/(2M)+U^\text{(HF)}_{i}$.
Writing down the expression associated with a diagram follows these simple rules:
% \begin{marginnote}
% 	In Hartree-Fock MBPT, the $-U$ part of $H_\text{1}=V-U$ cancels 
% 	all diagrams involving single-vertex loops~\cite{szabo,Drischler:2017wtt,1829502}.
% \end{marginnote}
\begin{itemize}
	\item Each vertex gives a factor $V^{ij,ab}$, with $i$ and $j$ %($a$ and $b$) being the %hole (particle) %lines directed towards (away from) the vertex,
	are the lines directed toward the
vertex, and $a$ and $b$ are the lines directed away from the vertex.

	\item Downwards lines give factors of $f_i$ while upwards lines give $(1-f_i)$, corresponding to hole and particle excitations of the reference ground-state, respectively, and

	\item For adjacent vertices there is an energy denominator given by subtracting the energy of the reference ground-state from the excited state corresponding to the particle and hole lines that are crossed by a virtual horizontal line between the two vertices. %orthogonal to the line that connects the two vertices.
	
	%\item the overall symmetry factor \com{if it's really so ``straightforward'' we should explain it in a sentence or two} %which can be inferred in a straightforward way from the respective diagrammatic structure~\cite{szabo}.
\end{itemize}
Each diagram's overall factor can be inferred 
%by counting pairs of equivalent lines, unique hole lines, and fermion loops~\cite{szabo}.
from the diagrammatic structure as well~\cite{szabo}.
%For instance, the expressions for the three third-order diagrams, i.e., the particle-particle (pp) and hole-hole (hh) ladder diagrams 
%as well as the particle-hole (ph) ring diagram,
%are given by
For instance, the expression of the third-order particle-particle (pp) diagram
reads
\begin{align}
	\mathcal{E}_{3,pp}(\kF)&=  
	\parbox[c][30pt]{40pt}{\input{diagrams/3rd_order_pp}}   = 
	\frac{1}{8}\sum_{ijabcd}
	V^{ij,ab}V^{ab,cd}V^{cd,ij} f_{ij}\bar f_{abcd} \frac{1}{D_{ab,ij}D_{cd,ij}}\,.
	%\\ \nonumber 
	%\\
	%E_{3,hh}(\kF)&=  
	%\parbox[c][30pt]{40pt}{\input{diagrams/3rd_order_hh}}  = 
	%\frac{1}{8}\sum_{ijklab}
	%V^{ij,ab}V^{kl,ij}V^{ab,kl} n_{ijkl}\bar n_{ab} \frac{1}{D_{ab,ij}D_{ab,kl}}\,,
	%\\ \nonumber 
	%\\
	%E_{3,ph}(\kF)&=  
	%\parbox[c][30pt]{40pt}{\input{diagrams/3rd_order_ph}}  = 
	%\sum_{ijkabc}
	%V^{ij,ab}V^{kb,ic}V^{ac,jk} n_{ijk}\bar n_{abc} \frac{1}{D_{ab,ij}D_{ac,jk}}\,.
\end{align}
Finding all valid diagrams (and associated expressions) at a given MBPT order has been formalized using graph-theory methods~\cite{Arthuis:2018yoo}:
%Arthuis:2020tjz
in Hartree-Fock MBPT there are $(1,3,39,840,27300)$ diagrams at order $l=(2,3,4,5,6)$.
Together with the automated code generation for the efficient Monte Carlo integration of arbitrary MBPT diagrams developed in Ref.~\cite{Drischler:2017wtt}, this has led to a fully automated approach to MBPT calculations.
%
% \begin{marginnote}
%     Using graph-theory methods one finds that in Hartree-Fock MBPT there are $(1,3,39,840,27300)$
%     diagrams at the MBPT orders $l=(2,3,4,5,6)$.
% \end{marginnote}

%%%Brueckner theory discussion
In the traditional Brueckner (or $G$-matrix) approach~\cite{RevModPhys.39.719}, 
%,Lejeune:1986aqu
the pp ladder diagrams are resummed to all orders, motivated by the large high-momentum components of traditional NN potentials to which the pp ladders are particularly sensitive. 
%The motivation for this approach comes from the feature that many traditional NN potentials have large high-momentum components, and the pp ladders are particularly sensitive to this feature. 
The pp bubbles in these diagrams are even ultraviolet divergent if the potential is not sufficiently suppressed at high momenta.
For modern low-momentum potentials, however, the pp ladders no longer play a distinguished role in the many-body expansion, 
and explicit MBPT calculations at third and fourth order have shown that they are not enhanced compared to other diagrams at the same order~\cite{Drischler:2017wtt}.
%This indicates that for low-momentum interactions, MBPT (without partial resummations) may be regarded as more systematic than Brueckner theory.
%Although not for the nuclear EOS, 
Nevertheless, partial diagrammatic resummations are still pertinent for consistent calculations of in-medium single-particle properties and response functions as performed in the self-consistent Green's functions method % is an improved approach compared to Brueckner theory 
(for more details see Section~\ref{sec24}).
The consistent generalization of MBPT to finite temperatures ($T>0$) is a nontrivial issue. 
From the standard finite-$T$ perturbation series for the grand-canonical 
potential\footnote{Using the grand-canonical ensemble is required for the evaluation of quantum-statistical averages in the thermodynamic limit.}
%~\cite{Fetter} %$\Omega(T,\mu)$,
\begin{align} \label{eq:MBPT_Tfin}
	\Omega(T,\mu)\simeq \Omega_0(T,\mu)+\sum_{l=1}^L \Omega_l(T,\mu)\,,
\end{align}
the free energy density $\mathcal{F}(T,\mu)$ is obtained via the 
%usual 
thermodynamic relation
%\begin{align} 
	$\mathcal{F}(T,\mu) = \Omega(T,\mu) + \mu  \, n(T,\mu)$.
%\end{align}
Here, the density is given by
%\begin{align} \label{eq:rhoderiv}
%\rho(T,\mu) = - \frac{\partial \Omega(T,\mu)}{\partial \mu}\,.
$n(T,\mu) = - \partial \Omega(T,\mu)/\partial \mu$.
%\end{align}
%The issue is now that the $T\to 0$ limit of the relation between $F$ and $\rho$ determined from these relations, in general, does not match the relation between $E$ and $n$ obtained from zero-$T$ MBPT.
The issue is now that the relations between $(\mathcal{F},n)$ and $(\mathcal{E},n)$ obtained in finite- and zero-$T$ MBPT, respectively, do not match in the limit $T\to0$
(as discussed further below).
%, generally.
%
% \begin{marginnote}
%     The use of the grand-canonical ensemble is required for
%     the evaluation of quantum-statistical averages in the thermodynamic limit.%~\cite{Fetter}. 
% \end{marginnote}

Regarding this, we first consider the finite-$T$ expression for the second-order diagram, %. It is given by
\begin{align}
\Omega_{2}(T,\mu)=&
 \parbox[c][30pt]{40pt}{\input{diagrams/2nd_order}}  = 
-\frac{1}{4}\sum_{ijab}
V^{ij,ab}V^{ab,ij} f_{ij}\bar f_{ab}\, \mathcal{G}_2\,. \label{eq:omega_2}
\end{align} 
Equation~\eqref{eq:omega_2}  differs only slightly from $\mathcal{E}_2(\kF)$ in Eq.~\eqref{eq:E2kf}.
First, 
%the $f_i=f_i(T,\mu)$ are Fermi-Dirac distributions instead of step functions centered at the 
%Fermi energy $\varepsilon_{\kF}$.
%Second,
the energy denominator is replaced by % the term
$\mathcal{G}_2 =(1-\e^{- D_{ab,ij}/T})/(2D_{ab,ij})$.
The numerator in this expression vanishes at any zero of the denominator, \ie, there are no poles at finite $T$.
In the $T\to 0$ limit the integration regions corresponding to the two terms in the numerator of $\mathcal{G}_2$ separate into two equivalent parts (with integrable poles at the integral boundaries), \ie,
$\mathcal{G}_2 \to 1/D_{ab,ij}$ for $T\to 0$.
These features pertain for higher-order diagrams~\cite{PhysRevC.99.065811}.\footnote{The poles (at the integral boundary) at $T=0$ lead to nonanalyticities in the asymmetry dependence of the nuclear EOS, see Section~\ref{sec32}.}
The second difference at finite $T$ compared to $T=0$ is that the $f_i=f_i(T,\mu)$ are Fermi-Dirac distributions instead of step functions centered at the 
Fermi energy $\varepsilon_{\kF}$.
%The $T\to 0$ limit of the perturbative contributions $\Omega_l$ then reproduces the corresponding zero-$T$ contributions $\mathcal{E}_l$ (for $l\in\{1,2,3\}$, see below), except for one crucial difference: the distribution functions are evaluated at the \textit{true} chemical potential $\mu$ instead of the \textit{reference} Fermi energy $\varepsilon_{\kF}$.
%
% \begin{marginnote}
% The poles (at the integral boundary)
% at $T=0$ lead to
% nonanalyticities in the asymmetry dependence of the
% nuclear EOS, see Section~\ref{sec32}.
% %For $l\geqslant 4$ ($l\geqslant 2$ for $U=0$), the relation %$\Omega_l(0,\varepsilon_{\kF})=\mathcal{E}_l(\kF)$ 
% %is not valid, due to anomalous terms in $\Omega_l$, see below.
% \end{marginnote}

Similar to the free Fermi gas (\ie, MBPT with $U=0$ and $L=0$), for Hartree-Fock MBPT at $L=1$, 
the chemical potential $\mu$ at $T=0$ matches the reference Fermi energy $\varepsilon_{\kF}$, 
with $n(T,\mu)=\sum_i f_i(T,\mu)$~\cite{PhysRevC.99.065811}.
But these relations cease to be valid at higher orders due to 
higher-order contributions in the expression for $n(T,\mu)$.
Note that these contributions involve factors $\partial f_i/\partial\mu = f_i(1-f_i)/T$, 
which become $\delta(\varepsilon_i-\mu)$ at $T=0$ (so there is a nonvanishing contribution at $T=0$).
Contributions involving factors $f_i(1-f_i)/T$ are also present 
in certain perturbative contributions to $\Omega$, 
starting at fourth order for Hartree-Fock MBPT~\cite{PhysRevC.99.065811}. [For $U=0$, they appear already at second order.]
These contributions can be associated with the presence of additional so-called anomalous diagrams in finite-$T$ MBPT, see Refs.~\cite{Kohn:1960zz,PhysRevC.99.065811} for more details.
As evident from the discussion above (\ie, below Eq.~\eqref{eq:omega_2}), the $T\to0$ limit of the finite-$T$ expressions for normal (\ie, not anomalous) contributions $\Omega_l$ matches the corresponding zero-$T$ contributions $\mathcal{E}_l$, except that the reference Fermi energy is replaced by the (true) chemical potential.
%As noted above, the finite-$T$ expressions for normal (\ie, not anomalous) contributions match those
%of zero-$T$ MBPT, except that the reference Fermi energy is replaced by the chemical potential.
Therefore, a consistent finite-$T$ version of Hartree-Fock MBPT for $L\leqslant 3$ would be given by
\begin{equation} \label{eq:MBPT_TfinF}
	\mathcal{F}(T,\tilde\mu)\simeq \mathcal{F}_0(T,\tilde\mu)+\sum_{l=1}^L \mathcal{F}_l(T,\tilde\mu)\,,
\end{equation}
where $\mathcal{F}_l=\Omega_l$ (for $l=1,2,3$) and the auxiliary ``chemical potential'' $\tilde\mu$ is related to the density 
via $n(T,\tilde\mu)=\sum_i f_i(T,\tilde\mu)$, implying $\tilde\mu \to \varepsilon_{\kF}$ in the $T\to 0$ limit.\footnote{The true chemical potential is obtained from $\mathcal{F}(T,\tilde\mu)$ via $\mu=\partial \mathcal{F}/\partial n$.}

In the $U=0$ case, the method for constructing a finite-$T$ perturbation series of the form of Eq.~\eqref{eq:MBPT_TfinF}
for any $L$ is well known~\cite{Kohn:1960zz}: %,brout1 
one expands each contribution to $\mathcal{F}(T,\mu)$ about $\tilde\mu$ 
according to $\mu=\tilde\mu +\sum_{l=1}^L \mu_l(T,\tilde\mu)$ 
while neglecting all terms beyond the truncation order~$L$. 
[The terms $\mu_l$ are determined by the requirement that the truncated expansion of %Eq.~\eqref{eq:rhoderiv} 
$n(T,\mu)$
about $\tilde\mu$ reproduces $n(T,\tilde\mu)=\sum_i f_i(T,\tilde\mu)$.]
This process can also be applied to Hartree-Fock MBPT~\cite{1829502},
with the caveat that the single-particle potential 
%is taken 
has 
to be evaluated at $\tilde\mu$, \ie, no derivatives of $U^\text{(HF)}_{i}(T,\tilde\mu)$  in $\tilde\mu$ appear.
%For both, t
In both cases, $U=0$ and Hartree-Fock MBPT, 
the resulting perturbation series for the free energy reproduces zero-$T$ MBPT at each 
truncation order $L$, even though the terms $\mathcal{F}_l$
contain anomalous contributions for $l\geqslant 4$ ($l\geqslant 2$, for $U=0$). 
%and the fact that it results from a truncated expansion of the original grand-canonical series shows that the latter is not consistent with zero-$T$ MBPT.
%Because this results from a truncated expansion of the original grand-canonical series, 
%Hartree-Fock MBPT cannot be consistent with zero-$T$ MBPT.
The fact that Eq.~\eqref{eq:MBPT_TfinF}
results from a truncated re-expansion
shows explicitly that the original grand-canonical series 
is not consistent with zero-$T$ MBPT.
For general arguments why the free-energy series is expected to give improved results 
compared to grand-canonical MBPT, %are discussed in 
see Refs.~\cite{PhysRevC.99.065811,Wellenhofer:2014hya}.
%
% \begin{marginnote}
% The terms $\mu_l$ are determined by the requirement that the truncated expansion of %Eq.~\eqref{eq:rhoderiv} 
% $n(T,\mu)$
% about $\tilde\mu$ reproduces $n(T,\tilde\mu)=\sum_i f_i(T,\tilde\mu)$.
% \end{marginnote}

Altogether, MBPT as formulated in the free-energy series [Eq.~\eqref{eq:MBPT_TfinF}] provides a consistent framework for nuclear matter calculations at zero- and finite-temperature, where many-body uncertainties can be systematically assessed by increasing the truncation order $L$. Although the number of MBPT diagrams increases rapidly with $L$, the technologies recently developed for automated diagram generation and evaluation~\cite{Arthuis:2018yoo,Drischler:2017wtt} enable calculations at high-enough orders to probe in detail the many-body convergence for chiral low-momentum NN and 3N interactions. 
Furthermore, exploring MBPT with single-particle potentials beyond the Hartree-Fock level is an important task for future research. In particular, the single-particle potential $U$ can be chosen at each truncation order such that the grand-canonical and free-energy series are also equivalent for $L>1$~\cite{PhysRevC.99.065811}; for Hartree-Fock MBPT, they are only equivalent for $L=1$. First investigations of this order-by-order renormalization of the single-particle potential have shown that higher-order contributions to $U$ can have a significant effect on low-order MBPT results and the many-body convergence~\cite{PhysRevC.95.034326}.
\subsection{Other many-body methods}
\label{sec24}

The advances in \ChEFT and RG methods
have established MBPT as a central approach for studying the nuclear EOS at zero and finite temperature.
While MBPT is 
%the most extensively studied approach at present and therefore 
the focus of this review,
%%the focus of this review, % and most extensively studied approach, 
%also 
various other many-body methods %frameworks 
have been applied %in nuclear matter 
in initial nuclear matter studies
with chiral NN and 3N interactions.
%(see Refs.~\addcite{} for recent reviews). 
In particular, nonperturbative frameworks are important to %assess
benchmark
the MBPT convergence and probe aspects of many-body physics beyond the nuclear EOS.  
Below we will briefly %highlight
discuss
the \textit{self-consistent Green's functions (SCGF) approach} and \textit{Quantum Monte Carlo (QMC) methods}.
%Important 
Other
methods not discussed here for brevity are coupled-cluster~(CC) theory~\cite{Hagen_2014}, the in-medium SRG~\cite{Hergert:2020bxy}, %Hergert:2015awm}, 
and lattice EFT~\cite{PhysRevLett.125.192502}. %Only comparisons of results obtained in different frameworks can draw a detailed picture of the nuclear EOS.
%Future comparisons of nuclear-matter results from these different frameworks will provide new insights regarding the description of nuclear interactions as well as many-body theory in general.
%Only through systematic comparisons between different many-body frameworks a coherent picture of microscopic interactions and the nuclear EOS can be drawn.
%In the future, systematic 
Systematic comparisons between different many-body frameworks will provide a coherent picture of microscopic interactions and nuclear many-body properties.

The SCGF approach~\cite{Dickhoff:2004xx,Rios:2020oad} is based on the
self-consistent computation of in-medium propagators (or Green's functions) in Fourier
(Matsubara) space, corresponding to the 
to-all-orders resummation of
% a subclass of
some
perturbative contributions to the propagators.
%Neutron and symmetric nuclear matter calculations both at zero~\cite{PhysRevC.88.044302} and finite temperature~\cite{Carbone:2018kji,Carbone:2014mja} have been implemented using 
SCGF calculations of the nuclear EOS at zero and finite temperature~\cite{Carbone:2014mja,Carbone:2018kji} have been implemented using
the in-medium $T$-matrix approximation, where the
%both pp and hh 
ladder diagrams are resummed to all orders, 
%corresponding to the 
%in line with the
%which amounts to the
providing
a thermodynamically consistent generalization of Brueckner theory~\cite{Rios:2020oad}.
Furthermore, SCGF calculations have been used to benchmark the order-by-order convergence of %MBPT(3) 
MBPT (up to third order)
 in neutron matter~\cite{PhysRevC.94.054307}. 
%Within many-body uncertainties the two approaches were in remarkable agreement 
The energy per particle obtained in SCGF and MBPT was found to agree well for a range of unevolved chiral NN and 3N interactions up to \NNNLO. % for a range of nuclear potentials.
The SCGF approach 
%as the advantage that it 
allows for fully consistent computations of response functions and transport properties, which will be vital for comparisons with MBPT calculations of these quantities.
%Future comparisons between MBPT and SCGF in this regard will be important. %would be vital for progress in nuclear many-body theory.

QMC refers to a family of stochastic methods
that solve the many-body Schr{\"o}dinger equation 
%computationally
%in terms of 
through
%a repeated 
random sampling~\cite{Lynn:2019rdt}.
As such, QMC methods are truly nonperturbative and provide important benchmarks for many-body methods with basis expansions.
However,
apart from the fermion sign problem 
a caveat
%regarding this 
is that most QMC methods require 
local nuclear potentials to obtain low-variance results, restricting both the regularization scheme and the interaction operators that can be included in the \ChEFT expansion. 
QMC calculations with local chiral NN and 3N potentials up to \NNLO have been carried out in neutron matter~\cite{Tews:2015ufa,Lynn:2015jua} and recently also symmetric nuclear matter~\cite{Lonardoni:2019ypg}.
Because of Fierz invariance breaking,
the regulator artifacts %in these calculations 
%appear to be somewhat large 
are %(due to Fierz-invariance breaking) 
significantly larger than in MBPT calculations with nonlocal potentials. 
On the other hand, since QMC methods are not restricted to soft interactions, a much wider range of momentum cutoffs can be studied with QMC. 
Hence, QMC methods can provide important insights into the residual cutoff dependence of observables and the breakdown scale of \ChEFT at high densities.

%It should be noted that, in addition to the chiral EFT potential approach employed here, 
%other approaches to nuclear many-body physics have been employed. 
%First, the EFT power counting can be applied directly to obervables instead of the nuclear interactions.
%This approach has been used in perturbative nuclear matter calculations with~\cite{Lacour:2009ej}
%and without~\cite{Lutz:1999vc,Kaiser:2001jx,Fiorilla:2011sr} partially resummed contact interactions.
%The EFT potential approach has so far however been the method of choice, also because of its greater flexibility regarding 
%the use of different many-body frameworks.
%It is also not clear whether a perturbative EFT approach (with power counting for obervables) can reach the same density regime as the potential approach.
%In the following, we therefore discuss only results based on modern chiral EFT potentials.

\subsection{Implementing three-body forces}
\label{sec25}

Three-nucleon forces are crucial for understanding properties of finite nuclei and nuclear matter~\cite{Hebeler:2015hla}, such as drip lines along isotopic chains and nuclear saturation in SNM. Even though partial-wave decomposed matrix elements of chiral 3N forces have become available recently up to \NNNLO~\cite{Hebeler:2015wxa},
%with high-enough truncation to ensure partial-wave convergence
implementing 3N forces in many-body calculations remains computationally difficult and usually requires approximations~\cite{Hebeler:2020ocj}. The large uncertainties due to 3N forces, \eg, in the nuclear EOS at densities $n \gtrsim n_0$, emphasize the need for improving these approximations as well as developing novel chiral NN and 3N potentials in general.

Normal ordering allows one to include dominant 3N contributions in many-body frameworks using density-dependent effective two-body potentials~\cite{holt20}. Through Wick's theorem the general three-body Hamiltonian can be \textit{exactly} normal ordered with respect to a finite-density reference state (\eg, the Fermi sea of noninteracting nucleons or the Hartree-Fock ground state) instead of the free-space vacuum~\cite{Bogner:2009bt}. This shifts contributions from the three-body Hamiltonian operator to effective zero-body, one-body, and two-body operators plus a residual (reduced) three-body operator. A many-body framework built for NN interactions can then incorporate a density-dependent effective interaction $V_{\rm NN}^{\rm med}$ derived from $V_{\rm 3N}$ as $V_\text{NN} \to V_\text{NN} + \xi\, V_{\rm NN}^{\rm med}$. 
The combinatorial factor $\xi$ is determined by Wick's theorem and depends on the many-body calculation of interest. % (see, \eg, Appendix~B in Ref.~\cite{Drischler:2016cpy} for details).
The matrix elements of $V_{\rm NN}^{\rm med}$ are obtained 
by summing one particle over the occupied states in the reference state:
%In infinite nuclear matter, normal ordering 3N forces typically involves summing one particle over the occupied states of the noninteracting Fermi sea:
\begin{equation} \label{eq:Veff}
	\Braket{\vb{2'3'}|V_{\rm NN}^{\rm med}|\bs{23}} 
	 = \enspace \parbox[c][32pt]{45pt}{\input{diagrams/normal_ordering}} 
	 = \sum \limits_{\sigma_1 \tau_1} \int \frac{\dd \vb{k}_1}{(2 \pi)^3} \, f_{\vb{1}} \Braket{\vb{12'3'}|\bar V_{\rm{3N}}|\vb{123}}  \, ,
\end{equation}
with the shorthand notation $\ket{\bs{i}} = \ket{\vb{k}_i\sigma_i \tau_i}$, antisymmetrized 3N interactions $\bar V_{\rm 3N}$, and momentum distribution function of the reference state $f_{\vb{1}}$.
%
%begin{marginnote}
%	\entry{One-body term}{The normal-ordered one-body term is relevant for single-particle energies.} 
%{In Hartree-Fock MBPT on has $\xi = 1$. 
%However, the contribution of the normal-ordered one-body operator 
%at first-order MBPT amounts to changing to $\xi = 1/3$ in that case.
%Similarly, $\xi = 1/2$ for first-order single-particle energies.}
%	\entry{Symmetry factor}{In a Hartree-Fock spectrum, one finds $\zeta = 1/2$ for single-particle energies; $\zeta = 1/3$ for the Hartree-Fock energy; and $\zeta = 1$ beyond the Hartree-Fock level.}
%\end{marginnote}

In contrast to the (Galilean-invariant) NN potential, the effective two-body potential~\eqref{eq:Veff} depends on the center-of-mass momentum $\vb{P}$ of the two remaining particles. Hence, both potentials cannot be straightforwardly combined in a partial-wave basis and different approximations for the $\vb{P}$ dependence have been used to enable applications to nuclear matter. Under the assumption that $\vb{P} = 0$ first implementations evaluated Eq.~\eqref{eq:Veff} semi-analytically in symmetric nuclear matter and pure neutron matter starting from the \NNLO 3N interactions~\cite{holt09,hebeler10,Hebeler:2010xb}. Extensions to asymmetric nuclear matter and finite temperature have followed~\cite{Wellenhofer:2014hya, Wellenhofer:2015qba, PhysRevC.94.054307, drischler16}, and a new method that allows for the construction of an effective two-body potential from any partial-wave decomposed 3N interaction in an improved  $\vb{P}$  angle-averaging approximation has been developed~\cite{drischler16}. The latter approach is especially advantageous for studying 3N forces at \NNNLO~\cite{drischler16}, bare and SRG-evolved, and in different regularization schemes. Semi-analytic expressions along the lines of Ref.~\cite{holt09} have been derived up to \NkLO{3}%~\cite{kaiser18,kaiser19,Kaiser:2020kpn} 
and also partially 
to \NkLO{4}~\cite{Kaiser:2020kpn}.
%~\cite{Kaiser:2019jvc,Kaiser:2020kpn}.
%Effective two-body potentials%~\eqref{eq:Veff} 
%have been applied %played an important role in studying 3N contributions to the nuclear EOS beyond Hartree-Fock~\addcite{}, BCS pairing gaps~\addcite{}, and optical potentials~\addcite{} up to \NNNLO. In particular, the partial-wave approach to normal ordering allowed for the first benchmark of MBPT(3) against the SCGF method to assess many-body convergence in PNM  with chiral NN and 3N forces at \NNNLO.

The three-body term in the normal-ordered Hamiltonian cannot be implemented using effective two-body potentials. In nuclear matter such residual 3N contributions have been studied in CC~\cite{Hagen:2013yba} and MBPT calculations~\cite{Kaiser:2012mm, Dyhdalo:2016ygz, Dyhdalo:2017gyl, Drischler:2017wtt}. Explicit calculations of the residual 3N diagram in MBPT at second order,
\begin{align}
	\mathcal{E}_{2}^\text{res}(\kF)=&
	\parbox[c][30pt]{40pt}{\input{diagrams/res_2nd_order}} = 
	-\frac{1}{36}\sum_{ijkabc}
	V^{ijk,abc}V^{abc,ijk} f_{ijk}\bar f_{abc} \frac{1}{D_{abc,ijk}}\,, \label{eq:E2res}
\end{align} 
showed for a range of chiral interactions that its contribution is typically much smaller than both the overall EFT truncation error and the individual contributions from the other MBPT diagrams up to this order~\cite{Drischler:2017wtt}. While these findings give some justification for the commonly used approximation where residual 3N contributions are neglected,
the automated approach introduced in Ref.~\cite{Drischler:2017wtt} implements chiral NN, 3N, and 4N interactions \textit{exactly} in nuclear matter calculations using a single-particle spin-isospin basis. % (see also Sec.~\ref{sec23}). %, \ie, without applying partial-wave decompositions. 
%Each matrix element is a function of the single-particle momenta, derived from the expressions of the nuclear interactions. 
Combined with high-performance computing, this method sets the stage for systematic studies of \ChEFT interactions in MBPT up to high orders without the mentioned approximations. 
%
% \begin{marginnote}
% 	\entry{Residual 3N diagram at second order}{\\ \\
% 		\input{diagrams/res_2nd_order}}
% \end{marginnote}

%% file: diagrams/2nd_order
\begin{fmffile}{diag_2nd_order}
	\begin{fmfgraph*}(30,40)
		\fmfcmd{style_def half_prop expr p =
			draw_plain p;
			shrink(.7);
			cfill (marrow (p, .5))
			endshrink;
			enddef;
		}
		\fmftop{v1}\fmfbottom{v0}
		\fmfv{d.shape=circle,d.filled=full,d.size=2thick}{v0}
		\fmfv{d.shape=circle,d.filled=full,d.size=2thick}{v1}
		\fmffreeze
		\fmf{half_prop,right=0.5}{v1,v0}
		\fmf{half_prop,left=0.5}{v1,v0}
		\fmf{half_prop,right=0.25}{v0,v1}
		\fmf{half_prop,left=0.25}{v0,v1}
	\end{fmfgraph*}
\end{fmffile}

%% file: diagrams/3rd_order_pp
\begin{fmffile}{diag_3rd_order_pp}
	\begin{fmfgraph*}(30,40)
		\fmfcmd{style_def half_prop expr p =
			draw_plain p;
			shrink(.7);
			cfill (marrow (p, .5))
			endshrink;
			enddef;
		}
		\fmftop{v2}\fmfbottom{v0}
		\fmfv{d.shape=circle,d.filled=full,d.size=2thick}{v0}
		\fmfv{d.shape=circle,d.filled=full,d.size=2thick}{v1}
		\fmfv{d.shape=circle,d.filled=full,d.size=2thick}{v2}
		\fmffreeze
		\fmf{half_prop,right=0.5}{v0,v1}
		\fmf{half_prop,left=0.5}{v0,v1}
		\fmf{half_prop,right=0.5}{v1,v2}
		\fmf{half_prop,left=0.5}{v1,v2}
		\fmf{half_prop,right=0.5}{v2,v0}
		\fmf{half_prop,left=0.5}{v2,v0}
	\end{fmfgraph*}
\end{fmffile}

%% file: diagrams/normal_ordering
\begin{fmffile}{diag_normal_ordering}
	\begin{fmfgraph*}(40,40)
		\fmfstraight
		\fmfbottom{i1,i2} 
		\fmftop{o1,o2}
		\fmffreeze
		\fmf{fermion}{i1,v1,o1}
		\fmf{fermion}{i2,v1,o2}	
		\fmf{fermion, tension=0.7,left=180}{v1,v1}		
		\fmfv{decor.shape=circle,decor.filled=empty,decor.size=20,label=$\bar V_{\rm 3N}$,label.dist=-1.3}{v1} 
		% "dist=-1.3" because the Ann. Rev. template moves LaTeX labels for unknown reason; unit mismatch?
	\end{fmfgraph*}
\end{fmffile}

%% file: diagrams/res_2nd_order
\begin{fmffile}{diag_res_2nd_order}
	\begin{fmfgraph*}(30,40)
		\fmfcmd{style_def half_prop expr p =
			draw_plain p;
			shrink(.7);
			cfill (marrow (p, .5))
			endshrink;
			enddef;
		}
		\fmftop{v1}\fmfbottom{v0}
		\fmfv{d.shape=circle,d.filled=full,d.size=2thick}{v0}
		\fmfv{d.shape=circle,d.filled=full,d.size=2thick}{v1}
		\fmffreeze
		\fmf{half_prop,right=0.5}{v1,v0}
		\fmf{half_prop,left=0.5}{v1,v0}
		\fmf{half_prop,left=0.75}{v1,v0}
		\fmf{half_prop,right=0.25}{v0,v1}
		\fmf{half_prop,left=0.25}{v0,v1}
		\fmf{half_prop,left=0.75}{v0,v1}
	\end{fmfgraph*}
\end{fmffile}

%% file: sections/03_nuclear_matter.tex
\section{Nuclear equation of state at zero and finite temperature} \label{sec:nuclear_matter}

In this Section we survey recent nuclear matter calculations up to $n\approx 2n_0$ in MBPT with chiral NN and 3N interactions. We discuss advances in the quantification and propagation of EFT truncation errors, confront different microscopic constraints on the nuclear symmetry energy with experiment, and examine contributions beyond the standard quadratic expansion of the EOS in the isospin asymmetry. We conclude the Section with results for the nuclear liquid-gas phase transition at finite temperature.

%In this section we survey recent applications of MBPT to the nuclear EOS based on \ChEFT interactions and discuss advances in the quantification of EFT truncation errors, which are the dominant theoretical uncertainties nowadays. In particular, we confront theory predictions for the nuclear symmetry with experimental constraints, and discuss the expansion in the (isospin-)asymmetry $\delta=(n_n-n_p)/n$ at higher orders. We conclude the section with results at finite temperature and nuclear thermodynamics, especially the liquid-gas phase transition.

\subsection{Confronting nuclear forces with empirical constraints} 
\label{sec:31}

%%%%%%%%%%%%%%%%%%%%%%%%%%%%%%%%%%%%%%%%
\begin{figure}[tb]
	\begin{center}
		\includegraphics[width=\textwidth]{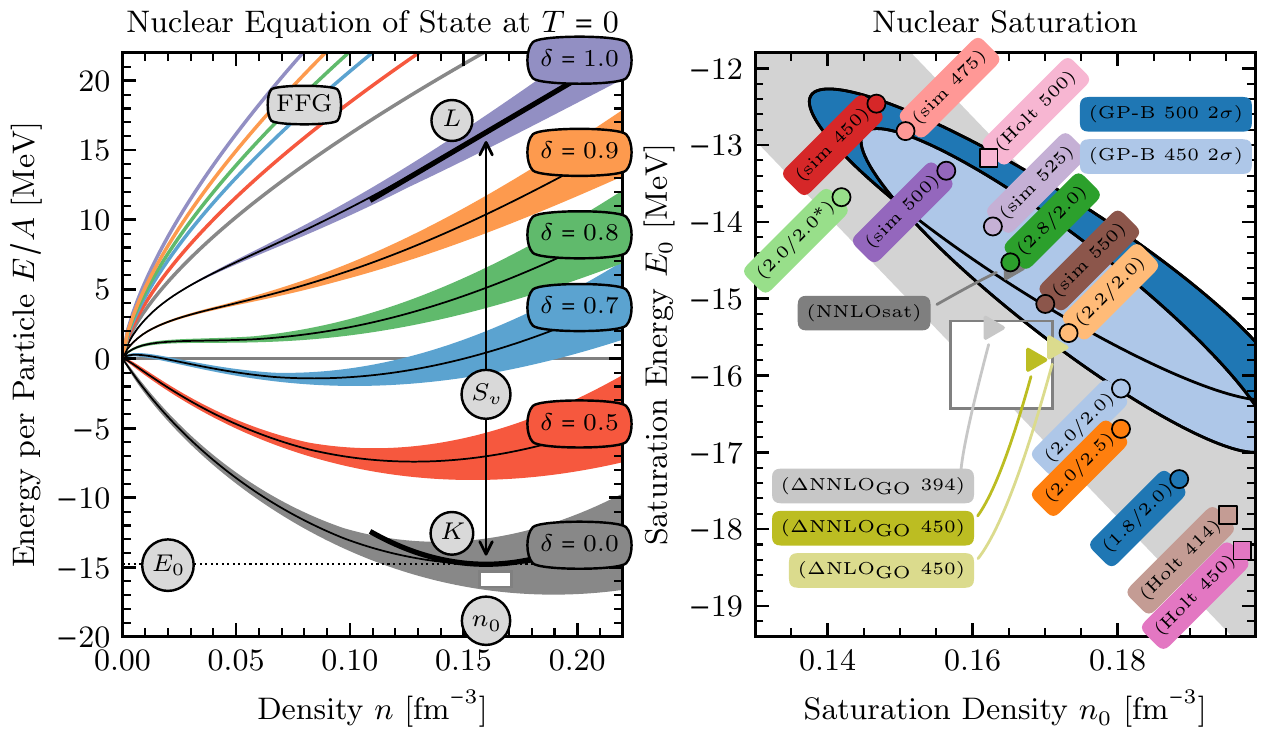}
	\end{center}
	\caption[ANM+Coester]{%
		(left)~Nuclear EOS at $T=0$ as a function of density $n$ for a representative set of isospin asymmetries $\delta$. %as obtained in Ref.~\cite{drischler16}. 
		The uncertainty bands in the energy per particle $E/A$ were obtained in Ref.~\cite{drischler16} by second-order MBPT calculations based on the Hebeler~\etal interactions~\cite{Hebeler:2010xb}.
		Key observables that characterize $E(n\approx n_0,\delta)/A$ are illustrated: the saturation point $(n_0, E_0)$, incompressibility $K_0$, symmetry energy $S_v$ and its slope parameter $L$ at $n_0$.
		(right)~Saturation points of numerous chiral interactions from fourth- (circles) and third-order (squares) MBPT calculations, as well as CC theory (triangles); specifically, the NN and 3N interactions by Hebeler~\etal~\cite{Hebeler:2010xb} (as in~(left), ``$\lambda/\Lambda_{\rm 3N} \, [\fmi]$''), Carlsson~\etal~\cite{Carlsson:2015vda} (``sim~$\Lambda \, [\MeV]$''), and Holt~\etal~\cite{PhysRevC.95.034326,Sammarruca:2014zia} (``Holt~$\Lambda \, [\MeV]$''). The ellipses show the $2\sigma$ regions of order-by-order calculations up to \NNNLO in MBPT with EFT truncation errors fully quantified~\cite{Drischler:2020yad}. Note that the saturation points are aligned along a Coester-like band (gray anticorrelation band).
		The white box in each panel depicts the empirical saturation point, $E_0 = -15.86 \pm 0.57 \MeV$ with $n_0 = 0.164 \pm 0.007 \fmiq$~\cite{drischler16}. The Hamiltonian ``2.0/2.0*’’ uses the $c_i$ values from the NN partial-wave analysis in Ref.~\cite{Rentmeester:2003mf} in the 3N forces.
		The data in the right panel
		were taken from Refs.~\cite{Jiang:2020the,Drischler:2017wtt,PhysRevC.95.034326,Drischler:2020yad}.
		%See the main text for details. 
	}
		\label{fig:anm_coester}
\end{figure}
%%%%%%%%%%%%%%%%%%%%%%%%%%%%%%%%%%%%%%%%

Figure~\ref{fig:anm_coester}~(left) illustrates the nuclear EOS at zero temperature as a function of density~$n$ for a representative set of isospin asymmetries $\delta=(n_n-n_p)/n$, where $n_n$  ($n_p$) is the neutron (proton) number density. 
%The uncertainty bands in the energy per particle $E/A$ were obtained in Ref.~\cite{drischler16} by second-order MBPT calculations based on the Hebeler~\etal interactions~\cite{Hebeler:2010xb}. 
Several general observations can be gleaned. Nuclear interactions are much stronger in SNM compared to PNM, which is closer to the free Fermi gas (FFG, solid lines). 
%While the uncertainties are thus larger in the vicinity of SNM, especially at densities $n \gtrsim n_0$, many-body calculations in PNM are typically well-controlled at $n \lesssim n_0$, and similar results can be obtained from a wide range of chiral NN and 3N interactions (see, \eg, Ref.~\cite{Huth:2020ozf}). 
Consequently, the uncertainties are larger in SNM, especially for densities $n \gtrsim n_0$.
In PNM they are well controlled for $n \lesssim n_0$, and a wide range of chiral NN and 3N interactions lead to similar results for PNM (see, \eg, Refs.~\cite{Coraggio13,Hebeler:2015hla,Huth:2020ozf}). Increasing uncertainties toward higher densities are predominantly due to 3N interactions.
Although the complexity of 3N interactions is much 
reduced in PNM~\cite{hebeler10}, they 
%which 
provide at all values of $\delta$ important repulsive contributions that grow stronger with the density than those of NN interactions. The 3N interactions are therefore crucial for understanding the high-density EOS and the structure of neutron stars. In PNM all chiral interactions up to \NNNLO are completely determined by the $\pi$N and NN system. The intermediate- and short-range 3N interactions at \NNLO that are proportional to the LECs $c_D$ and $c_E$, respectively, vanish (for regulators symmetric in the particle labels) due to the coupling of pions to spin and the Pauli principle, respectively. Also the long-range two-pion exchange 3N forces at \NNLO are simplified since the LEC $c_4$ does not contribute. This allows for tight low-density constraints on the neutron-rich matter EOS from PNM calculations and systematic high-density extrapolations (see Section~\ref{sec:applications}).
\begin{marginnote}
	\entry{PNM}{pure neutron matter ($\delta=1$)}
	\entry{ANM}{asymmetric nuclear matter ($0<\delta<1$)}	
	\entry{SNM}{symmetric nuclear matter ($\delta=0$)}
	\entry{FFG}{free Fermi gas}
\end{marginnote}

Nuclear matter represents an ideal system for testing nuclear interactions at the densities accessible to laboratory experiments and their implementation in many-body methods. As illustrated in Fig.~\ref{fig:anm_coester}~(left), the nuclear EOS in the vicinity of $n_0$ is (to good approximation) characterized by only a few experimentally accessible quantities. That is, the EOS of SNM can be expanded about its minimum $n_0$ as $E(n,\delta=0)/A\approx E_0+(K/2) \, \eta^2$, with the saturation energy $E_0 = E(n_0,0)/A$, incompressibility~$K$, and $\eta= (n-n_0)/(3n_0)$. Further, explicit ANM calculations with chiral NN and 3N interactions%at $n \lesssim n_0$ 
~\cite{drischler16, Drischler:2013iza, Wellenhofer:2016lnl} have shown that the %isospin-
asymmetry dependence of the nuclear EOS is reasonably well reproduced by the standard quadratic approximation $E(n,\delta)/A=E(n,0)/A + E_\text{sym}(n) \, \delta^2$, where the %nuclear (deleted for space)
symmetry energy expanded in density reads $E_\text{sym}(n) \approx S_v + L\,\eta$. 
In this approximation one finds $E(n,1)/A\approx (E_0 + S_v) + L\,\eta$ for PNM. Microscopic predictions and empirical constraints for $(n_0,E_0,K)$ and $(S_v,L)$ can then be confronted with one another.\footnote{See Ref.~\cite{Roca-Maza:2018ujj} for a review of the link between the nuclear EOS and nuclear observables; \eg, from measurements of the isoscalar giant monopole resonance $K \approx 210-230 \MeV$ was inferred.} %, \eg, $K$ can be inferred from measurements of giant monopole resonances~\cite{Blaizot:1980tw}. 
% \begin{marginnote}
% 	\entry{CC}{coupled cluster (see Sec.~\ref{sec24})}
% \end{marginnote}

Nuclear saturation emerges from a delicate cancellation between kinetic and interaction contributions to the EOS. Reproducing empirical constraints on $(n_0,E_0,K)$ is therefore an important benchmark of nuclear interactions, especially 3N forces (providing the necessary repulsion). Figure~\ref{fig:anm_coester}~(right) %summarizes 
depicts
the saturation points of numerous chiral potentials as predicted by fourth- (circles) and third-order (squares) MBPT calculations. The saturation points are aligned along a Coester-like band (gray anticorrelation band), which overlaps with the empirical saturation point (white boxes, see legend), determined from a set of energy density functionals~\cite{Drischler:2016cpy}. 
%Despite overall reasonable saturation properties, 
%However, not a single deltaless chiral Hamiltonian shown in Fig.~\ref{fig:anm_coester} actually saturates inside the empirical range, including the CC result for \NNLOsat~($\textcolor{tab20:14}{\smallblacktriangleright}$). 
Also shown are results from CC calculations with \NNLOsat~($\textcolor{tab20:14}{\smallblacktriangleright}$) and the new deltaful chiral potentials~\cite{Jiang:2020the} ``$\Delta$NLO\textsubscript{GO} $\Lambda \, [\MeV]$'' and ``$\Delta$NNLO\textsubscript{GO} $\Lambda \, [\MeV]$'' ($\textcolor{tab20:15}{\smallblacktriangleright}$, $\textcolor{tab20:16}{\smallblacktriangleright}$, $\textcolor{tab20:17}{\smallblacktriangleright}$), constructed by the Gothenburg-Oak-Ridge~(GO) collaboration. Only the latter fall into the empirical range for $(n_0,E_0)$. However, judging the extent a nuclear potential reproduces empirical (saturation) properties can be quite misleading without taking meaningful uncertainties into account; especially, the truncation of the EFT expansion at a finite order can result in sizable EFT truncation errors (even at \NNNLO) that need to be quantified.
%
% \begin{marginnote}
% 	%The range of the incompressibility $K$ obtained from nuclear matter calculations with \ChEFT potentials is around $K = 215(40) \MeV$~\cite{Drischler:2017wtt,drischler16}.
% 	See Ref.~\cite{Roca-Maza:2018ujj} for a review of the link between the nuclear EOS and nuclear observables; \eg, from measurements of the isoscalar giant monopole resonance $K \approx 210-230 \MeV$ 
% 	was inferred.
% 	%can be inferred.
% \end{marginnote}

Until a few years ago, the prevalent way of estimating theoretical uncertainties in nuclear matter calculations was parameter variation within some (arbitrary) range; that is, probing the observable's sensitivity to, \eg, the 3N LECs or momentum cutoff. Recently, the focus has been more
%there has been a concerted effort to focus 
on the systematic 
quantification of EFT truncation errors~\cite{Furnstahl:2014xsa}, which can be estimated by
%because recent advances in the optimization of chiral interactions suggest that the statistical uncertainties in the LECs are much smaller.
assuming that an observable's EFT
%(oldversion) 
%%expansion also holds for individual observables at a given order~$k$, \ie, 
%(newversion1) expansion of the interactions at a given order~$k$ translates into an expansion for a given observable $y_k(n)$ of the form
%(newversion2) 
convergence pattern at order~$k$ 
%of a given observable $\genobs_k(\kinparvec)$
%computed from chiral interactions at orders $m=0,1,\ldots,k$
takes the form
$\genobs_k(\kinparvec) = \genobsref(\kinparvec) \sum_{m=0}^k c_m(\kinparvec) Q^m(\kinparvec)$~\cite{Melendez:2019izc}. Here, $\genobsref(\kinparvec)$  sets a dimensionful reference scale, $Q(\kinparvec)$ is the dimensionless expansion parameter, and the $c_m(\kinparvec)$ are the dimensionless coefficients not to be confused with the LECs of the interaction (\eg, $\genobs_4 =E/A$ at \NNNLO). Note that $c_1=0$ in Weinberg power counting. For given choices of $\genobsref(\kinparvec)$ and $Q(\kinparvec)$,\footnote{For instance, $\genobsref \simeq \genobs_0$ and $Q \sim p/\Lambda_b$ with the typical momentum $p \propto \kF(n)$ and some estimate of the EFT breakdown scale~$\Lambda_b$ have been used to estimate uncertainties in the nuclear EOS.} the $c_{m\leqslant k}(\kinparvec)$ are obtained from order-by-order calculations $\lbrace \genobs_0, \genobs_1, \ldots, \genobs_k\rbrace$ of the observable. Since $\genobsref(\kinparvec)$ and $Q(\kinparvec)$ factor in all physical scales, the $c_m(\kinparvec)$ are expected to be of order one (\ie, \textit{natural}), unless the coefficients are fine-tuned. The standard EFT uncertainty, which assumes that the truncation error  is dominated by the first omitted term, has been implemented by Epelbaum~\etal~\cite{Epelbaum:2014efa} and applied to a wide range of observables in finite nuclei and infinite matter. This ``EKM uncertainty'' can be summarized at \NkLO{j} as $\delta \genobs(\kinparvec) = \genobsref \, Q^{j+2} \max( |c_0|, |c_1|, \dots, |c_{j+1}|)$, 
%whose point-wise estimates are in semi-quantitative agreement with Bayesian analyses with certain prior choices for $c_m$~\cite{Furnstahl:2015rha}. 
whose point-wise estimates can be interpreted as Bayesian credibility regions under a particular choice of priors for $c_m$~\cite{Furnstahl:2015rha}. 
%Different modifications have been studied; \eg, to avoid issues with LO contributions.
%
% \begin{marginnote}
% 	Both $\genobsref(\kinparvec)$ and $Q(\kinparvec)$ need to be chosen; \eg, $\genobsref \simeq \genobs_0$ and $Q \sim p/\Lambda_b$ with typical momentum $p \propto \kF(n)$ and some estimate of the EFT breakdown scale~$\Lambda_b$ have been applied to the nuclear EOS.
% \end{marginnote}

The \textit{Bayesian Uncertainty Quantification: Errors in Your EFT} (BUQEYE) Collaboration has recently introduced a Bayesian framework for quantifying correlated EFT uncertainties in the nuclear EOS~\cite{Drischler:2020hwi, Drischler:2020yad}.\footnote{The framework is publicly available at \url{https://buqeye.github.io/software/}.} In contrast to the standard EFT uncertainty, the new framework allows for the quantification \textit{and} propagation of statistically meaningful uncertainties to derived quantities (\eg, the pressure) while accounting for correlations across densities and between observables. Without considering these correlations, uncertainties can be overestimated. The framework also includes Bayesian model checking tools~\cite{Melendez:2017phj} for diagnosing and testing whether the in-medium \ChEFT expansion works as assumed (\eg, inference for $\Lambda_b$).
Gaussian Processes (GPs) with physics-based hyperparameters are trained on the order-by-order calculations of the energy per particle under the assumption that all $c_m(\kinparvec)$ are 
%independent and identically distributed 
random curves drawn from a single GP~\cite{Melendez:2019izc}. The Gaussian posterior for the $c_m(\kinparvec)$ is then used to %derive the to-all-order EFT truncation error 
estimate the to-all-orders EFT truncation error
$\delta\genobs_k(\kinparvec) = \genobsref(\kinparvec) \sum_{m=k+1}^\infty c_m(\kinparvec) Q^m(\kinparvec)$ and combined with additional (\eg, many-body) uncertainties. From the posterior also arbitrary derivatives in $\kinparvec$ can be obtained.
\begin{marginnote}
	\entry{BUQEYE}{Bayesian Uncertainty Quantification: Errors in Your EFT}%; the collaboration's framework for quantifying correlated EFT truncation errors is publicly available at \url{buqeye.github.io/software/}.}
	%\entry{GP}{Gaussian Process}
	\entry{GP}{Gaussian Process}
	\entry{GP--B}{Gaussian Process--BUQEYE Collaboration}
\end{marginnote}

%%%%%%%%%%%%%%%%%%%%%%%%%%%%%%%%%%%%%%%%
\begin{figure}[tb]
	\begin{center}
		\includegraphics[width=\textwidth]{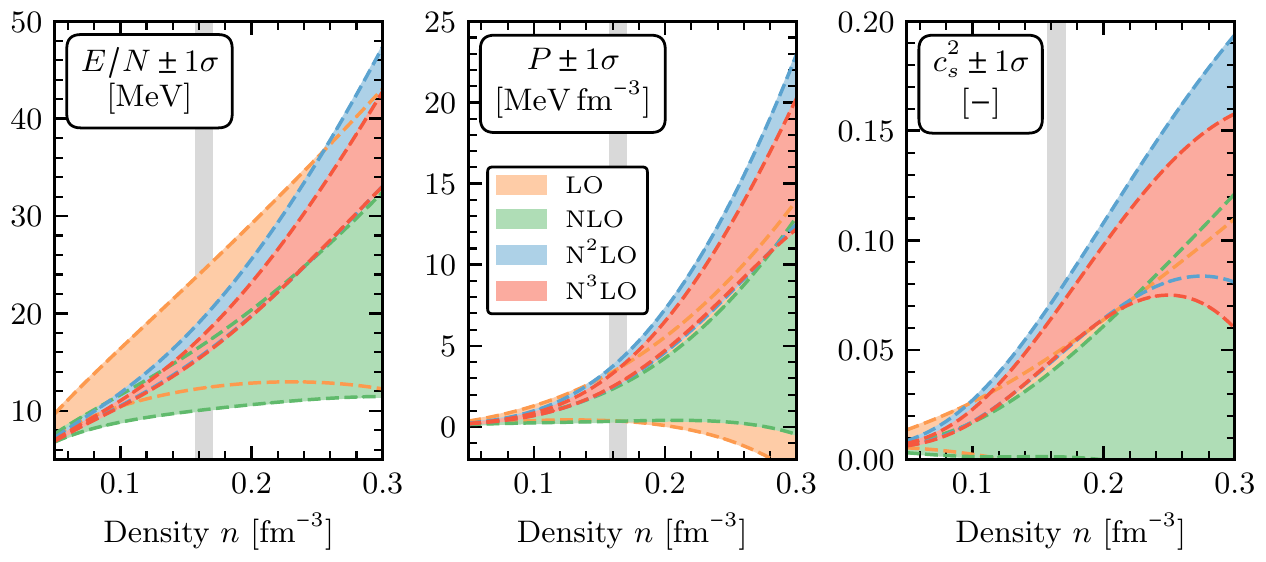}	
	\end{center}
	\caption[PNM EOS]{Order-by-order predictions for the energy per particle $E/N$ (left panel), pressure $P=n^2 \dd (E/N) \! / \dd n$ (middle panel), and speed of sound squared $c_s^2 =  \partial P/\partial \varepsilon$ (right panel) in PNM as a function of the density~\cite{Drischler:2020hwi} based on the MBPT calculations up to \NNNLO in Ref.~\cite{Drischler:2017wtt}. The energy density $\varepsilon = n (E/N + m_n)$ includes the neutron rest mass energy $m_n$. Correlated uncertainty bands are given at the $1\sigma$ confidence level. The data were taken from Ref.~\cite{Drischler:2020hwi}. %See the main text for details.
	}
	\label{fig:eft_conv}
\end{figure}
%%%%%%%%%%%%%%%%%%%%%%%%%%%%%%%%%%%%%%%%

Using this new framework, Drischler~\etal~\cite{Drischler:2020hwi,Drischler:2020yad} studied the EFT convergence of the first order-by-order calculations with NN and 3N interactions up to \NNNLO in PNM and SNM, conducted in Refs.~\cite{Drischler:2017wtt,Leonhardt:2019fua} using a novel Monte Carlo integration framework for MBPT. % with adequate many-body convergence. 
The associated \NNNLO 4N Hartree-Fock energies have been found negligible compared to the overall uncertainties (see also Ref.~\cite{PhysRevLett.110.032504}).
To construct a set of order-by-order NN and 3N interactions up to \NNNLO, Ref.~\cite{Drischler:2017wtt} adjusted the two 3N LECs to the triton and $(n_0,E_0)$ for two cutoffs. Several potentials with reasonable saturation properties were obtained, although generally underbound at \NNNLO. This holds also at the $2\sigma$ credibility level with EFT truncation errors quantified~\cite{Drischler:2020yad}, as depicted by the ellipses ``\GPB{$\Lambda$ [MeV]}'' in Fig.~\ref{fig:anm_coester}~(right). %Hoppe~\etal~\cite{Hoppe:2019uyw} found that the corresponding binding energies (charge radii) of medium-mass nuclei are predicted too small (too large) compared to experiment, in agreement (disagreement) with the expectations from SNM. 
Hoppe~\etal~\cite{Hoppe:2019uyw} found that, in agreement with the expectations from SNM, predicted values for the corresponding binding energies of medium-mass nuclei are too small compared with experiment and, in contrast to the SNM expectations, predicted values for the corresponding charge radii are too large.
Since both observables were also much more sensitive to the 3N LEC~$c_D$, SNM and medium-mass nuclei seem more intricately connected than one might naively expect~\cite{Hoppe:2019uyw}.

Figure~\ref{fig:eft_conv} shows the order-by-order predictions for the energy per particle, pressure, and speed of sound squared in PNM at the $1\sigma$ confidence level based on ``\GPB{500}''~\cite{Drischler:2020hwi}. The observables show an order-by-order convergence pattern at $n \lesssim 0.1\fmiq$, whereas \NNLO and \NNNLO have a markedly different density dependence at $n \gtrsim n_0$ due to repulsive 3N contributions. This is also manifested in the Bayesian diagnostics~\cite{Melendez:2019izc}. Assuming $Q=\kF/\Lambda_b$, the inferred breakdown scale $\Lambda_b \approx 600 \MeV$ is consistent with free-space NN scattering and could be associated with $n > 2n_0$. The EFT truncation errors are strongly correlated in density and to those in SNM. A correlated approach is therefore necessary to propagate uncertainties reliably to derived quantities, although the standard EFT uncertainty for the energy per particle is broadly similar to the $1\sigma$ confidence level~\cite{Drischler:2020hwi}. % .%, \ie, the energy per particle

%One can then interpolate between PNM and SNM to obtain neutron-star matter (\ie, matter in beta-equilibrium), which is with $\delta \approx 0.9$ very neutron-rich, but not exactly PNM. There exist only a few explicit calculations of ANM, mostly in MBPT because the isospin-asymmetry can be varied relatively straightforwardly (for results in the SCGF method see Ref.~\addcite{}). 

\subsection{Nuclear symmetry energy and the isospin-asymmetry expansion}
\label{sec32}

%%%%%%%%%%%%%%%%%%%%%%%%%%%%%%%%%%%%%%%%
\begin{figure}[tb]
	\begin{center}
		\includegraphics[width=\textwidth]{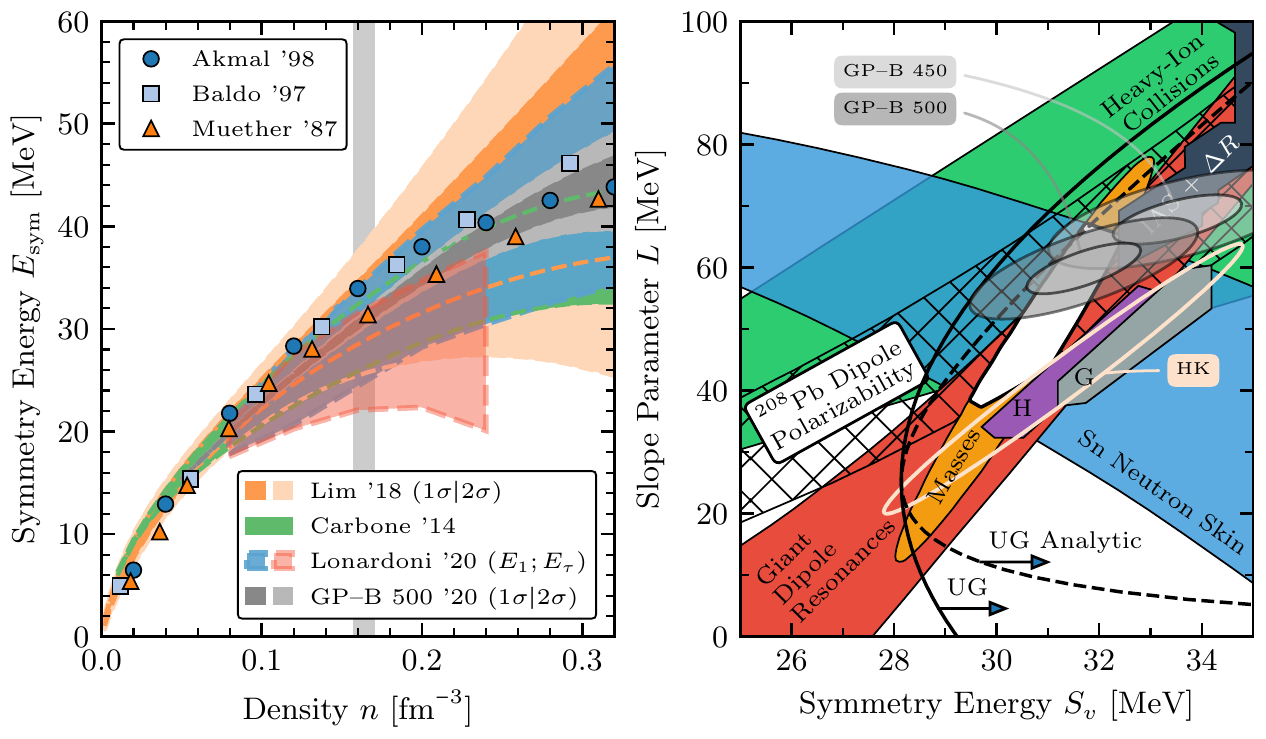}
	\end{center}
	\caption[Symmetry energy]{
		(left)~Constraints on $E_\text{sym}(n)$ based on chiral interactions (bands) and phenomenological potentials (symbols). The vertical band depicts the empirical saturation density. Dashed lines are used to enhance readability. Data were taken from Akmal~\etal~\cite{akmal98}, Baldo~\etal~\cite{baldo97},
Muether~\etal~\cite{muether87}, Lim~\& Holt~\cite{lim18},
Carbone~\etal~\cite{Carbone:2014mja}, Lonardoni~\etal~\cite{Lonardoni:2019ypg},
and Drischler~\etal~\cite{Drischler:2020yad, Drischler:2017wtt} [``\GPB{500}''].
		(right)~Theoretical and experimental constraints for $(S_v, L)$ as well as the conjectured UG bounds~\cite{Tews17NMatter} in comparison (see annotations in the panel). The experimental constraints are derived from heavy-ion collisions (HIC)~\cite{Tsan09Lconstr}, neutron-skin thicknesses of \isotope{Sn} isotopes~\cite{Chen10skin}, giant dipole resonances (GDR)~\cite{Trip08gdr}, the dipole polarizability of \isotope[208]{Pb}~\cite{Tami11dipole,Roca13dipoPb},
  nuclear masses~\cite{Kort10edf}, and isobaric analog states
 and isovector skins ($\text{IAS}+\Delta R$)~\cite{Dani17IVskins}. In addition,
 theoretical constraints derived from microscopic neutron-matter calculations by
 Hebeler~\etal~(H)~\cite{hebeler10prl}, Holt \& Kaiser~(HK)~\cite{PhysRevC.95.034326}, and \mbox{Gandolfi}~\etal~(G)~\cite{Gandolfi:2011xu}.
		Gray ellipses~\cite{Drischler:2020yad} show the allowed regions from PNM and SNM calculations at \NNNLO with truncation errors quantified (light: $1\sigma$, dark: $2\sigma$).
		%The joint experimental constraint (white area) only barely overlaps with the analysis of isobaric analog states and isovector skins ($\text{IAS}+\Delta R$). 
		The white area in the center shows the joint experimental constraint;
the constraints extracted from measurements of $\text{IAS}+\Delta R$ are not included in this
area because they barely overlap.% without ``$\text{IAS}+\Delta R$''.
		%For more details see Refs.~\addcite{} and the main text.
	}
	\label{fig:esym}
\end{figure}
%%%%%%%%%%%%%%%%%%%%%%%%%%%%%%%%%%%%%%%%

The nuclear symmetry energy %, which in 
%the standard quadratic approximation is given by $E_\text{sym}(n) = E(n,1)/A - E(n,0)/A$, %
is a key quantity to understand the structure of neutron-rich nuclei and neutron stars. Although masses of heavy nuclei constrain the value of the symmetry energy well at nuclear densities, its density dependence is much less known~\cite{Horowitz:2014bja}. Studying the density-dependent symmetry energy from theory, experiment, and observation is therefore an important task in the era of multimessenger astronomy.

Figure~\ref{fig:esym}~(left) summarizes theoretical constraints for $E_\text{sym}(n\leqslant 2\nsat)$ from a wide range of chiral NN and 3N forces as well as different many-body methods. Specifically, we show the results for $E_\text{sym}(n) = E(n,\delta=1)/A-E(n,0)/A$ as obtained in the calculations by Lim~\& Holt~\cite{lim18} and Drischler~\etal~\cite{Drischler:2020yad, Drischler:2017wtt} [``\GPB{500}''] in MBPT, Carbone~\etal~\cite{Carbone:2014mja} in the SCGF method, and Lonardoni~\etal~\cite{Lonardoni:2019ypg} using QMC methods. The latter were conducted with two different parameterizations of the \NNLO 3N contact interaction (\ie, distinct bands for $E_1$ and $E_\tau$) to demonstrate the significant regulator artifacts present in local chiral 3N potentials. Different uncertainty estimates were used in these calculations. The uncertainty bands by Carbone~\etal probe parameter variations in the nuclear interactions, while those by Lonardoni~\etal and Drischler~\etal quantify truncation errors using the standard EFT uncertainty (up to \NNLO) and BUQEYE's new Bayesian framework (up to \NNNLO), respectively. Also many-body (or statistical Monte Carlo) uncertainties are included in the bands. Lim~\& Holt performed a statistical analysis of MBPT calculations based on a range of chiral potentials at different orders and two single-particle spectra to probe the chiral and many-body convergence. Only the results by Lim~\& Holt and Drischler~\etal (both MBPT) have a clear statistical interpretation, each at the $1\sigma$ and $2\sigma$ confidence level (different shadings). Overall, the constraints from \ChEFT are consistent with each other, even at the highest densities shown, but the uncertainties in $E_\text{sym}(n)$ are generally sizable, \eg, $20.7 \pm 1.1$, $31.5 \pm 3.0$, and $49.0 \pm 12.0 \MeV$ at $n_0/2$, $n_0$, and $2\nsat$, respectively, for Lim~\& Holt at the $1\sigma$ confidence level. Drawing general conclusions from comparing the sizes of these bands can be misleading since the underlying methods for estimating uncertainties are quite different. Order-by-order comparisons for a wider range of chiral NN and 3N interactions with EFT truncation errors quantified are called for to provide more insights in and stringent constraints on $E_\text{sym}(n)$. The Bayesian statistical tools introduced by the BUQEYE Collaboration allow for such systematic studies.%\footnote{Note that the statistical meaning of the (uncorrelated) standard EFT uncertainty applied to $E_\text{sym}(n)$ is unclear and requires a choice for the typical momentum $p$ that depends on two different $\kF$ at fixed density $n$.}

Despite the large uncertainties in the SNM EOS (see Sec.~\ref{sec:31}), predictions for  $E_\text{sym}(n)$ [as an energy difference] can be made with significantly smaller uncertainties than those in $E(n,1)/A$ and $E(n,0)/A$ individually, if correlations are properly considered. 
The BUQEYE Collaboration~\cite{Drischler:2020yad} found that the EFT truncation errors associated with the PNM and SNM calculations in Ref.~\cite{Drischler:2017wtt} are \textit{highly correlated}, meaning that the uncertainty in $E_\text{sym}(n)$ is less than the usual in-quadrature sum of errors. Combined with order-by-order calculations up to \NNNLO this led to the narrow constraints [see gray bands in
%bands ``\GPB{500}'' in 
Fig.~\ref{fig:esym}~(left)] based on the interactions used with $\Lambda = 500 \MeV$ (\eg, $E_\text{sym}(2n_0) = 45.0 \pm 2.8 \MeV$). Another set with $\Lambda = 450 \MeV$ is compatible at the $2\sigma$ confidence level. The bands agree with the constraints by Lim~\& Holt at the $1\sigma$ level (or even better) as well as the calculations by Akmal~\etal~\cite{akmal98}, Baldo~\etal~\cite{baldo97}, and Muether~\etal~\cite{muether87} with phenomenological nuclear potentials. The latter, however, do not provide uncertainties that could be used to judge the level of agreement. These correlations need to be investigated further using different many-body frameworks and interactions.
%
% \begin{marginnote}
% 	The statistical meaning of the (uncorrelated) standard EFT uncertainty applied to $E_\text{sym}(n)$ is unclear and requires a choice for the typical momentum $p$ that depends on two different $\kF$ at fixed density $n$.
% \end{marginnote}

%With the uncertainties in the symmetry energy coming mainly from SNM, improved predictions for the symmetry parameters near saturation density, ($S_v$,$L$),  can be obtained from microscopic PNM calculations by using the empirical information on the saturation point.
Figure~\ref{fig:esym}~(right) compares various theoretical and experimental constraints in the $S_v$--$L$ plane (see annotations). 
%; i.e., from heavy-ion collisions (HIC), neutron-skin thicknesses of \isotope{Sn} isotopes, giant dipole resonances (GDR), the dipole polarizability of \isotope[208]{Pb}, and nuclear masses. 
The %allowed 
regions obtained by Hebeler~\etal~\cite[``H'']{hebeler10prl}, Gandolfi~\etal\cite[``G'']{Gandolfi:2011xu}, and Holt \& Kaiser~\cite[``HK'']{PhysRevC.95.034326}, which were derived from microscopic PNM calculations and the empirical saturation point, %properties, %only, 
%are in remarkable with each other and
agree well with each other and are consistent with the range in $S_v$ of the joint experimental constraint (white area), although $L$ is predicted with somewhat lower values. 
%Constraints extracted from measurements of isobaric analog states and isovector skins (``$\text{IAS}+\Delta R$'') are not included in the white area because they barely overlap.
The $1\sigma$ and $2\sigma$ ellipses of ``\GPB{500}'' (as in Fig.~\ref{fig:esym}~(left)) are in excellent agreement with the joint experimental constraint [``\GPB{450}'' is slightly shifted to higher $(S_v,L)$], indicating a stiffer neutron-rich matter EOS at $n_0$ compared to the other theoretical constraints. This is, however, consistent with joint theory-agnostic posteriors from pulsar, gravitational-wave, and NICER observations (\eg, compare with Figure~1 in Ref.~\cite{Essick:2020flb}). An important feature of the correlated GP approach is that the theoretical uncertainties in $n_0$ (including truncation errors) are accounted for through marginalization over the Gaussian posterior for the saturation density predicted from the SNM calculations, $\pr(n_0) \approx 0.17 \pm 0.01 \fmiq$. 
%An important feature of the correlated GP approach is that the uncertainties in $n_0$ are accounted for through marginalizing instead of a point estimate, \ie, $\pr(S_v,L \given \mathcal{D}) = \int \pr(E_\text{sym}, L \given n_0, \mathcal{D}) \pr(n_0 \given \mathcal{D}) \dd{n_0}$, where $\pr(E_\text{sym}, L \given n_0, \mathcal{D})$ denotes the correlated to-all-orders posterior at a particular saturation density $n_0$, and $\pr(n_0 \given \mathcal{D}) \approx 0.17 \pm 0.01 \fmiq$ is the Gaussian posterior for the saturation density (including truncation errors) predicted from the SNM calculations. 
Apart from the calculations by Holt \& Kaiser (``HK'') allowing slightly lower $(S_v,L)$, all shown theory calculations satisfy the constraint (solid black line) derived from the conjecture~\cite{Tews17NMatter} that the unitary gas (UG) sets a lower bound for the PNM EOS. [Note that Ref.~\cite{Tews17NMatter} also made additional assumptions to derive an analytic bound (dashed black line).]
%Overall, Fig.~\ref{fig:esym}~(right) shows that the uncertainties in $S_v$ and especially in $L$ are significant, both in nuclear theory and experiment. Within these uncertainties,
%different predictions are consistent with experimental constraints and for $S_v$ also with each other.
Overall, Fig.~\ref{fig:esym}~(right) shows that current constraints from nuclear theory and experiment predict the symmetry energy parameters in the range %$S_v\approx 30-34 \MeV$ and $L\approx 35-65 \MeV$.
$S_v\approx 28-35 \MeV$ and  $L\approx 20-72 \MeV$.
\begin{marginnote}
	\entry{UG}{unitary gas}%; Ref.~\cite{Tews17NMatter} also made additional  assumptions to derive an analytic bound (``UG Analytic'', dashed black line in Fig.~\ref{fig:esym}~(right)). %, which agrees only at the $1\sigma$ level with the ``GP--B'' constraints.}
\end{marginnote}

%modern theory predictions of nuclear bulk properties agree (within uncertainties) already reasonably well with experimental determinations of these quantities. 

%%%%%%
%Preliminary results from the \isotope{Pb} Radius Experiment (PREX)~II using parity-violating electron scattering seem to confirm the large neutron skins in \isotope[208]{Pb} found by PREX~I, corresponding to high values for $L$ and thus a stiff nuclear EOS at nuclear densities. \com{PREX is not exactly the best example for strong experimental constraints on the EOS. Nevertheless pretty recent. Can we replace it with something else?}
%%%%%%%%%%%%%%

While the standard quadratic approximation [$E(n,\delta)/A=E(n,0)/A + E_\text{sym}(n) \, \delta^2$] is in general sufficient to characterize the isospin-asymmetry dependence of the nuclear EOS, certain
%Contributions to the isospin-asymmetry dependence of the nuclear EOS beyond the standard quadratic approximation affect 
neutron-star properties, 
%the density-dependent proton fraction
%at beta equilibrium, 
such as the crust-core transition~\cite{Cai:2011zn} and the threshold for the direct URCA cooling process~\cite{Steiner:2006bx}, are sensitive to nonquadratic contributions. Neglecting charge-symmetry breaking effects, the energy per particle 
%has commonly been
may be assumed to have an expansion in the asymmetry $\delta$ of the form $E(n, \delta)/A \approx  E(n,0)/A + \sum_{l=1}^L S_{2l}(n)\, \delta^{2l}$, %Here, t
where the standard quadratic approximation corresponds to %keeping only the leading $S_{2}(n)$ term, \ie,
%$E_\text{sym}(n) = S_{2}(n)$  and 
$S_{2l>2}(n)=0$.
%, but 
Note, however, that already the FFG contributes to the nonquadratic terms, \eg, $S_4^\text{FFG}(n) \simeq 0.45 \MeV \times \left( n/n_0 \right)^{2/3}$. Parametric fits to microscopic ANM calculations have been used to investigate the nonquadratic contributions and found them to be relatively small~\cite{drischler16, Drischler:2013iza, Somasundaram:2020chb, Wellenhofer:2016lnl}.
Recently, however, Kaiser~\cite{Kaiser:2015qia} has shown that MBPT at second order gives rise to additional logarithmic contributions $\sim \delta^{2l}\ln|\delta|$ with $l\geqslant 2$.
%Related to this, 
%At finite temperature,
%Ref.~\cite{Wellenhofer:2016lnl} 
Furthermore, Wellenhofer~\etal~\cite{Wellenhofer:2016lnl} found that the analogous expansion of the free energy exhibits convergent behavior 
%(up to $\delta=1$) 
for $\delta \leqslant 1$
only at high temperature. 
That is, 
%the convergence radius of the Taylor expansion about $\delta=0$ 
the expansion's radius of convergence
decreases to zero in the limit $T \rightarrow 0$ (with diverging
$S_{2l>2}$), as implied by the logarithmic terms at $T=0$.
Nevertheless, Wen \& Holt~\cite{wen20} demonstrated that the coefficients of the normal and logarithmic terms at $T=0$ can be extracted up to ${\cal O}(\delta^6)$ from high-precision MBPT calculations with chiral interactions. Such calculations allow for the improvement of existing parametrizations in $\delta$ at $T=0$ and help motivate the investigation of alternative schemes, such as an expansion in terms of the proton fraction 
 $x = n_p/n = (1-\delta)/2$
%(instead of $\delta = 1-2x$) 
%about $\delta = 1-2x = 1$ 
for neutron-rich matter.

\subsection{Nuclear thermodynamics}
\label{sec33}

%%%%%%%%%%%%%%%%%%%%%%%%%%%%%%%%%%%%%%%%
\begin{figure*}[tb]
	\begin{center}
		\includegraphics[width=\textwidth]{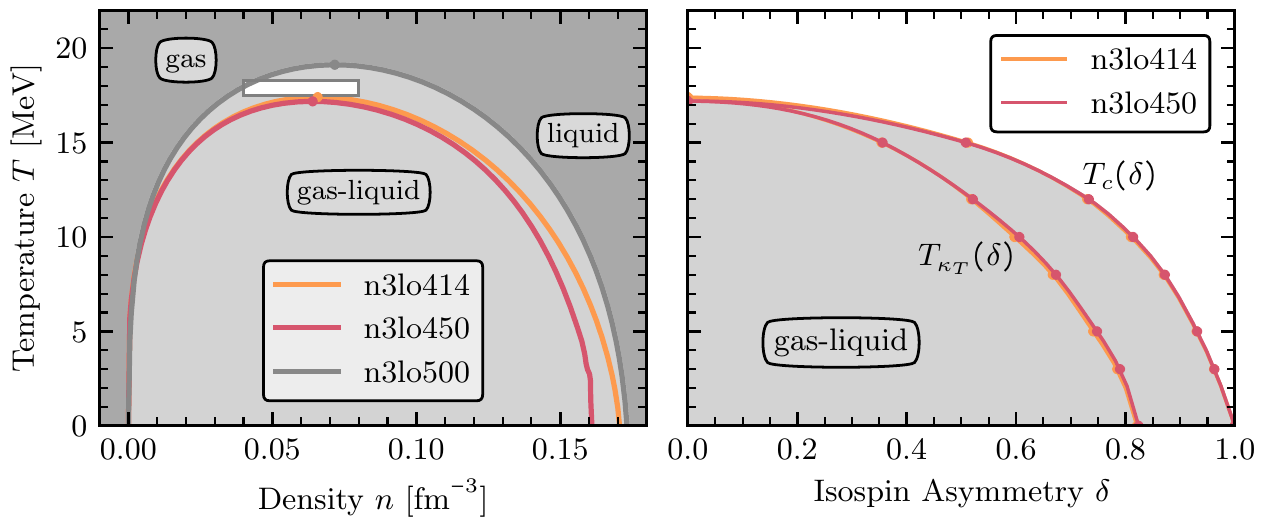}
	\end{center}
	\caption[Finite $T$ EOS]{
		(left)~Liquid-gas coexistence boundary (binodal) of SNM from second-order MBPT calculations based on 
		three 
		%different 
		sets of \NNNLO NN potentials and \NNLO 3N interactions with $\Lambda=414$, $450$, and $500\MeV$~\cite{Wellenhofer:2014hya,Wellenhofer:2015qba}. The zero-temperature limit of the coexistence boundary corresponds to the nuclear saturation point.
		The white box shows the empirical range for the critical point from Ref.~\cite{Elliott:2013pna}. 
		%inferred from multifragmentation experiments in Ref.~\cite{Elliott:2013pna}.
		%obtained from multifragmentation experiments~\cite{Elliott:2013pna}.
		%from Ref.~\cite{Elliott:2013pna}.
		(right)~Asymmetry dependence of the critical temperature $T_c(\delta)$ and
		%Also shown is the asymmetry dependence of 
		the temperature $T_{\kappa_T}(\delta)$ where the region with negative
		$\kappa_T^{-1}$ vanishes.
		%The results were obtained in second-order MBPT based on \NNNLO NN potentials and \NNLO 3N interactions with $\Lambda=414$, $450$, and $500\MeV$~\cite{Wellenhofer:2014hya,Wellenhofer:2015qba}.
	}
	\label{fig:liquidgas}
\end{figure*}
%%%%%%%%%%%%%%%%%%%%%%%%%%%%%%%%%%%%%%%%

While thermal effects are negligible in isolated neutron stars, they become important in neutron star mergers and core-collapse supernovae, where 
%temperatures 
$T \lesssim 100 \MeV$ can be reached. Dense matter at such high temperatures not only consists of nucleons and leptons but also of additional particles such as pions and hyperons. The consistent inclusion of these particles in medium is work in %and not further discussed in this review article 
%(see Refs.~\cite{Fore:2019wib,Petschauer:2020urh} for recent work). 
progress~\cite{Fore:2019wib,Petschauer:2020urh}.
In the nascent field of multimessenger astronomy, one of the immediate theoretical needs is consistent modeling of (i) cold neutron stars, (ii) hot hypermassive neutron stars formed in the aftermath of compact object mergers, and (iii) core-collapse supernovae so that observations and simulations in any one of these astrophysical regimes can be propagated to constrain the others. Finite-temperature MBPT with \ChEFT interactions is %emerging as 
a suitable framework for this purpose, and here we describe some of the 
%recent progress in describing nuclear thermodynamics from 
results on nuclear thermodynamics in recent years
(for reviews, see Refs.~\cite{Holt:2013fwa,Holt:2014hma}).

The salient thermodynamic feature of 
homogeneous 
nuclear matter at sub-saturation densities is the presence of a liquid-gas type instability
%of homogeneous matter with respect to
toward the formation of clustered structures.
In neutron stars, this instability corresponds to the 
crust-core
transition, % from the crust to the inner core, 
involving 
such intricate features as a variety of ``pasta'' shapes~\cite{PhysRevC.88.065807}.
The nuclear liquid-gas instability is also connected to the observed multifragmentation events in intermediate-energy heavy-ion collisions.
In the idealized case of (infinite) 
%homogeneous 
nuclear matter, there is a liquid-gas phase transition of van-der-Waals type.
Nuclear matter calculations at finite temperature with chiral interactions 
have provided predictions for the properties of this phase transition, in particular the location of the critical point.
Figure~\ref{fig:liquidgas}~(left) shows 
%the 
the second-order MBPT results for the boundary of the liquid-gas coexistence region (so-called binodal) of SNM
obtained in Ref.~\cite{Wellenhofer:2014hya}.
%[The binodal of SNM was recently computed using the SCGF method~\cite{Carbone:2018kji} and with lattice EFT~\cite{PhysRevLett.125.192502}; the results are similar to those in Fig.~\ref{fig:liquidgas}.]
[The results for the binodal of SNM recently obtained with the SCGF method~\cite{Carbone:2018kji} and lattice EFT~\cite{PhysRevLett.125.192502} are similar to those in Fig.~\ref{fig:liquidgas}.]
The 
%results for the 
predicted critical point, especially the associated temperature $T_c\approx 17-19 \MeV$, 
is consistent with estimates (\eg, $T_c\approx 15-20 \MeV$~\cite{Karnaukhov:2008be}) extracted from multifragmentation, nuclear fission, and
compound nuclear decay experiments~\cite{Karnaukhov:2008be,Elliott:2013pna}.
%
% \begin{marginnote}
% The binodal of SNM was recently computed using the SCGF method~\cite{Carbone:2018kji} and with lattice EFT~\cite{PhysRevLett.125.192502}; the results are similar to those of Fig.~\ref{fig:liquidgas}. %are similar to the MBPT results discussed here.
% \end{marginnote}

In the interior of the binodal 
%there is 
a region where the homogeneous system is unstable with respect to 
infinitesimal density fluctuations 
can be found. 
The boundary of this region is called spinodal. Between the binodal and spinodal the uniform system is metastable.
[The two boundaries coincide at the critical point.]
%whereas at the critical point the two coincide.
% the binodal and spinodal coincide.
For SNM, %the interior of the spinodal corresponds to 
the unstable region 
%the region where the 
is identified by a negative
inverse isothermal compressibility, $\kappa_T^{-1}=n(\partial P/\partial n)<0$.
%is negative.
An equivalent stability criterion is $\partial \mu/\partial n>0$, 
%which is equivalent to the free energy $F(T,n)$ being a strictly convex function of $n$. 
corresponding to a strictly convex free energy density $\mathcal{F}(T,n)$ as a function of $n$.
If charge-symmetry breaking effects are neglected, SNM 
can be treated as
a pure substance with one particle species (nucleons),
%a single-species substance, 
whereas
%, but 
ANM is a binary mixture with two thermodynamically distinct particles (neutrons and protons).
This implies that the stability criteria are different in the two cases,
%This implies that the stability criteria for ANM are different than those for SNM.
%In particular, 
and for ANM the region with $\kappa_T^{-1}<0$ is a subregion of the spinodal region.
There are various equivalent stability criteria for binary mixtures~\cite{beegle}.
A useful criterion is that outside the spinodal the free energy $\mathcal{F}(T,n_{n},n_{p})$ is a strictly convex function of 
$n_{n}$ and $n_{p}$ (see Ref.~\cite{Wellenhofer:2015qba} for details).
The MBPT results for the asymmetry dependence of the critical temperature $T_c(\delta)$ from Ref.~\cite{Wellenhofer:2015qba} 
are shown in Fig.~\ref{fig:liquidgas}~(right).
For comparison, we also show the trajectory of the temperature
$T_{\kappa_T}(\delta)$ where the subregion with negative
$\kappa_T^{-1}$ vanishes.
The trajectory of $T_c(\delta)$ reaches its $T=0$ endpoint at a small %value of the 
proton fraction~$x$; \ie, while PNM is stable at all densities, already 
%a small proton fraction 
small $x$ lead to a region where the system undergoes a phase separation~\cite{Wellenhofer:2015qba,Ducoin:2005aa}.
%
%\begin{marginnote}
%While for SNM the binodal corresponds to the Maxwell construction,
%for ANM the more involved Gibbs construction is required~\cite{Ducoin:2005aa}.
%\end{marginnote}
%\begin{marginnote}
%The different properties of the liquid-gas instability in SNM and ANM imply that
%nuclear matter can be classified as an azeotrope, with the special feature that the proton fraction
%at the azeotropic point is (by charge symmetry) constrained to the value $x=0.5$.
%\end{marginnote}

A useful characteristic for the temperature dependence of the 
%EOS of homogeneous nuclear matter 
nuclear EOS
is the thermal index $\Gamma_\text{th}(T,n,\delta)=
1+P_\text{th}(T,n,\delta)/\mathcal{E}_\text{th}(T,n,\delta)$, 
where $P_\text{th}$ is the thermal part of the pressure, and $\mathcal{E}_\text{th}$ is the thermal energy density.
For a free gas of nucleons with effective masses $m^*_{n,p}(n,\delta)$
one obtains for $\Gamma_\text{th}$ the following temperature-independent expression\footnote{The thermal index of a free Fermi gas is $\Gamma_\text{th,free}=5/3$.}
\begin{align}\label{eq:Gth_m*ANMclass}
    \Gamma_\text{th}^\star(n,\delta)
    &= \frac{5}{3}-
    \sum\limits_{t=\text{n},\text{p}}
    \frac{n_{t}(n,\delta)}{m^*_t(n,\delta)}\frac{\partial m^*_t(n,\delta)}{\partial n}.
\end{align}
[To be precise, for $\delta\notin\{0,1\}$ the above expression is valid only in the classical limit, 
but it provides a good approximation to $\Gamma_\text{th}^\star(n,x)$ for intermediate values of $\delta$ as Ref.~\cite{Huth:2020ozf} showed.]
Recently, Refs.~\cite{1829502,Carbone:2019pkr}
showed that $\Gamma_\text{th}^\star(n,\delta)$ reproduces 
%the exact thermal index $\Gamma_\text{th}(T,\rho,\delta)$ 
the exact $\Gamma_\text{th}$
with high accuracy.
This implies that the temperature-dependence of the EOS can be characterized in terms 
of a temperature-independent effective mass (see Ref.~\cite{Huth:2020ozf} for a recent implementation), which is in particular useful for monitoring thermal effects in astrophysical applications~\cite{yasin20,Bauswein:2010dn}. 
%\com{Should end on page 17, give or take.}

%\begin{afterthoughts}[Afterthoughts: remove prior to submission]
%	\begin{enumerate}
%		\item More details about Arianna's results? Especially most recently with Achim?
%		\item Related but applies to all sections: as explained in the Introduction the review focuses on MBPT for several good reasons. Since we three have sort of ``dominated'' the field of nuclear EOS calculations using MBPT it's only natural that lots of our papers are cited. Nevertheless, we need to make sure that also our colleagues get credit for their works.
%	\end{enumerate}
%\end{afterthoughts}

%% file: sections/04_applications.tex
\section{Applications to neutron star physics} \label{sec:applications}

In this Section our goal is to emphasize the prominent role of nuclear theory in modeling neutron stars, core-collapse supernovae, and neutron star mergers. We begin by placing high-energy nuclear astrophysics in the more general context of the QCD phase diagram and discuss under what ambient conditions \ChEFT can serve as a tool to constrain the properties of hot and dense matter. Specific applications include the neutron-star mass-radius relation, moment of inertia, and tidal deformability, as well as the nuclear EOS and neutrino opacity for astrophysical simulations.

\subsection{Scales in hot and dense stellar matter}

The extreme astrophysical environments found in core-collapse supernovae, neutron star interiors, and neutron-star mergers span baryon number densities $n_B \sim 10^{-7} - 10^1\,n_0$, temperatures $T \sim 0 - 100\MeV$, and isospin asymmetries $\delta \sim 0-1$ (corresponding to electron fractions $Y_e \sim 0-0.5$) \cite{oertel17}. In Sections~\ref{sec:mb_problem} and \ref{sec:nuclear_matter} we have shown that \ChEFT provides a suitable framework to constrain the EOS, transport, and response properties of hadronic matter when the physical energy scale is well below the chiral symmetry breaking scale of $\Lambda_\chi \sim 1\GeV$. In practice, \ChEFT descriptions of nuclear matter based on high-precision NN and 3N forces begin to break down at densities $n \approx 2-3\,n_0$ and temperatures $T \lesssim 30\MeV$. Therefore, additional modeling is needed at high densities and temperatures to cover all regions of astrophysical interest. For this purpose, high-energy heavy-ion collisions at RHIC, LHC, and especially FAIR aim to probe states of matter similar to those that exist naturally in neutron stars, but reaching sufficiently large proton-neutron asymmetries remains a significant challenge that may be addressed with next-generation radioactive ion beam facilities, such as FRIB. The interplay of microscopic \ChEFT, whose convergence pattern is not especially sensitive to the isospin asymmetry, together with upcoming nuclear experiments that create and study hot, dense, and neutron-rich matter, will provide a direct line of inquiry probing neutron-star physics from low to high densities.

\begin{figure}[tb]
		\begin{centering}
			\includegraphics[width=0.83\textwidth]{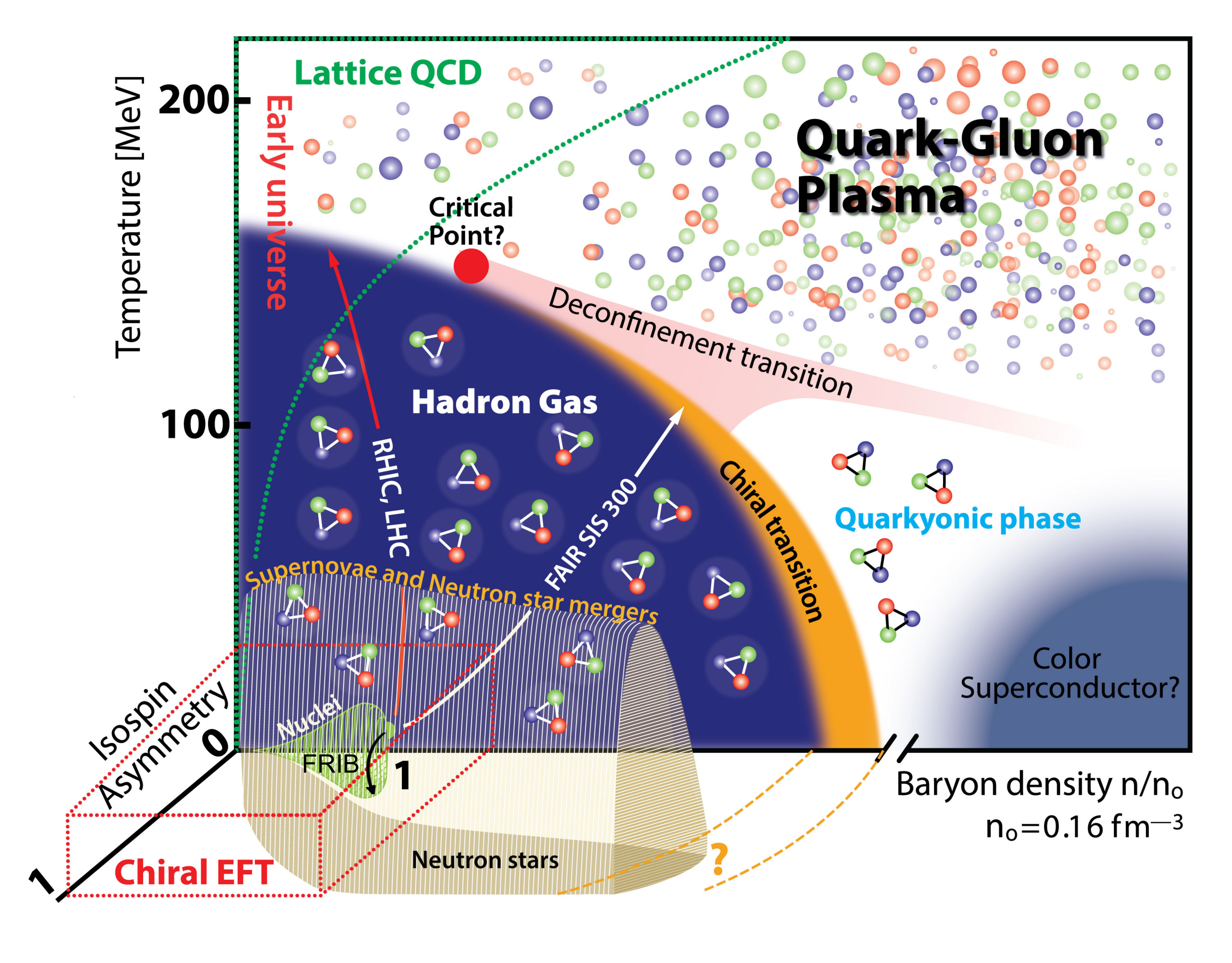}
		\end{centering}
		\caption[QCD phase diagram]{Schematic view of the QCD phase diagram. We highlight regions probed by experiments (RHIC, LHC, FAIR, and FRIB), regions of validity for lattice QCD and \ChEFT, and environments reached in neutron stars, supernovae, and neutron star mergers.}
		\label{fig:phase}
	\end{figure}

From the observational side, measurements of neutron star masses, radii, tidal deformabilities, and moments of inertia are expected to place constraints on the pressure of beta-equilibrium matter at $n \gtrsim 2n_0$~\cite{lattimer01,lim18,tsang19}. In Fig.\ \ref{fig:phase}, we present a qualitative overview of the QCD phase diagram and highlight regions probed by nuclear experiments (RHIC, LHC, FAIR, and FRIB), theory (lattice QCD and \ChEFT), and astrophysical simulations of neutron stars, supernovae, and neutron star mergers. We see that \ChEFT intersects strongly with the region of FRIB experiments and nuclear astrophysics, providing a bridge between new discoveries in the laboratory and their implications for neutron stars. The next decade is expected to witness a strong interplay among all of these different fields,
%as fundamental nuclear physics is 
with nuclear theory predictions getting
confronted with stringent empirical tests.% from astrophysical observations.
\begin{marginnote}
	\entry{RHIC}{Relativistic Heavy Ion Collider}
	\entry{LHC}{Large Hadron Collider}
	\entry{FAIR}{Facility for Antiproton and Ion Research}
	\entry{FRIB}{Facility for Rare Isotope Beams}
\end{marginnote}

%%%%%%%%%%%%%%%%%%%%%%%%%%%%%%%%%%%%%%%%%%%%%%%%%%%%%%%%%%%%%%%%%%%%%%%%%%%%%%%%

%%%%%%%%%%%%%%%%%%%%%%%%%%%%%%%%%%%%%%%%%%%%%%%%%%%%%%%%%%%%%%%%%%%%%%%%%%%%%%%%

\subsection{Neutron star structure}
The mass-radius relation of non-rotating neutron stars is determined from the EOS by the general relativistic equations for hydrostatic equilibrium, the Tolmann-Oppenheimer-Volkoff (TOV) equations:
\begin{equation}
\dv{p}{r} = - \frac{G(M(r)+4\pi r^{3} p)(\varepsilon + p)}
{r(r-2GM(r))}, \hspace{.3in} \dv{M}{r} = 4\pi r^{2}\varepsilon,
\label{eq:tov}
\end{equation}
where $r$ is the radial distance from the center of the star, $M(r)$ is the mass enclosed within $r$, $\varepsilon$ is the energy density, and $p$ is the pressure. Analysis of spectral data from neutron stars in quiescent low-mass x-ray binaries and x-ray bursters \cite{ozel16,nattil17} have resulted in radius measurements $R_{1.5} = 10 - 13\km$ for typical $1.5\Msun$ neutron stars. More recently, the NICER x-ray telescope has observed hot spot emissions from the accretion-powered x-ray pulsar PSR J0030+045. Pulse profile modeling of the x-ray spectrum from two independent groups have yielded consistent results for the neutron star's mass $M = 1.44^{+0.15}_{-0.14}\Msun$ \cite{Miller_2019} and $M = 1.34^{+0.15}_{-0.16}\Msun$ \cite{Riley_2019} and radius $R = 13.02^{+1.24}_{-1.06}\km$~\cite{Miller_2019} and $R = 12.71^{+1.14}_{-1.19}\km$~\cite{Riley_2019} at the 68\% credibility level. Future large area x-ray timing instruments, such as \mbox{STROBE-X} %~\cite{strobex}
and eXTP, %~\cite{extp}
have the potential to reduce uncertainties in the neutron-star mass-radius relation to $\sim 2\%$ at a given value of the mass. This would significantly constrain the neutron-rich matter EOS at $n\approx 2n_0$ and when combined with mass and radius measurements of the heaviest neutron stars could give hints about the composition of the inner core~\cite{annala19}.
\begin{marginnote}
	\entry{TOV}{Tolmann-Oppenheimer-Volkoff}
    \entry{NICER}{Neutron star Interior Composition ExploreR}
    \entry{STROBE-X}{Spectroscopic Time-Resolving Observatory for Broadband Energy X-rays}
    \entry{eXTP}{enhanced X-ray Timing and Polarimetry}
\end{marginnote}

\begin{figure}[tb]
        \begin{center} \includegraphics[width=0.60\textwidth]{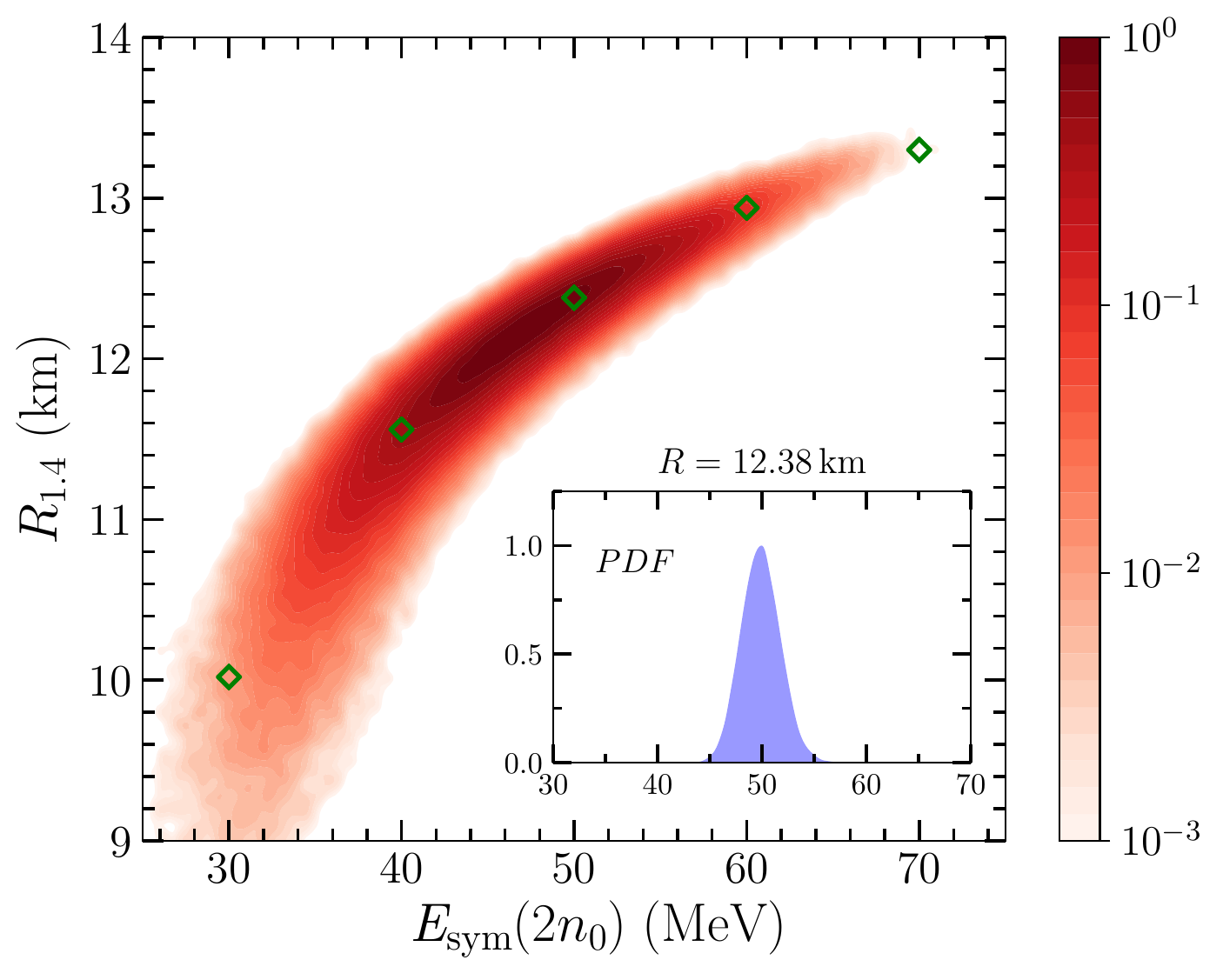}
        \end{center}
        \caption{Correlations between the radius of a $1.4\Msun$ neutron star and the isospin asymmetry energy $E_{\rm sym}$ at twice saturation density $n=2n_0$. The inset shows the probability distribution function (PDF) of $E_{\rm sym}(2n_0)$ for the specific value $R_{1.4}=12.38\km$. Results obtained from the Bayesian modeling of the nuclear EOS in Ref.\ \cite{lim18}.
        }
        \label{fig:R14_S2}
\end{figure}

In recent years numerous works have studied constraints on the neutron star EOS from \ChEFT.\footnote{In addition to a high-density extrapolation, 
a uniform-matter EOS from \ChEFT needs to be
supplemented with a neutron-star crust model, \eg, the BPS crust model~\cite{Baym:1971pw}.}
In Ref.\ \cite{hebeler10prl} the EOS of neutron-rich matter was calculated up to saturation density 
with MBPT using chiral NN and 3N interactions.
To extrapolate to higher densities, a series of piecewise polytropes was used to parameterize the EOS. It was found that \ChEFT generically gives rise to soft EOSs that lead to $1.4\Msun$ neutron stars with radii in the range $R_{1.4}=10-14 \km$. Subsequent studies (\eg, Refs.~\cite{Greif:2018njt,tews19epj,lim19epj,Huth:2020ozf}) have employed a wider range of chiral forces, increased the assumed range of validity for \ChEFT calculations to $2n_0$, and explored other high-density EOSs, including smooth extrapolations and speed of sound parameterizations. The choice of transition density at which \ChEFT predictions are replaced by model-dependent high-density parameterizations has a particularly large influence on neutron-star radius constraints. For instance, when the transition density was raised to $n_t=2n_0$, Ref.\ \cite{tews18} obtained $R_{1.4}=9.4-12.3\km$ while Ref.\ \cite{lim19epj} found $R_{1.4}=10.3-12.9\km$, both calculations eliminating the stiffest EOSs that would give rise to $R_{1.4}>13\km$. To demonstrate how a precise neutron-star mass and radius measurement can constrain the EOS of beta-equilibrium matter at $n=2n_0$, in Fig.\ \ref{fig:R14_S2} we show the correlated probability distribution~\cite{lim18} for the radius of a $1.4\Msun$ neutron star and the nuclear symmetry energy at twice saturation density $E_{\rm sym}(2n_0)$. In the inset we show the conditional probability distribution for $E_{\rm sym}(2n_0)$ assuming a precise measurement of $R_{1.4} = 12.38\km$. % and note that $E_{\rm sym}(2n_0)$ can be determined to within approximately 10\% for the EOS modeling~\cite{lim18} used in the analysis.
For the specific EOS modeling used in Ref.~\cite{lim18},
such a precise radius constraint determines $E_{\rm sym}(2n_0)$ with an uncertainty of approximately 10\%. 
%
%\begin{marginnote}
%In addition to a high-density extrapolation, a uniform-matter EOS from \ChEFT needs to be supplemented with a neutron-star crust model, \eg, the BPS crust model~\cite{Baym:1971pw}.
%\entry{BPS}{Baym, Pethick, and Sutherland}
%\end{marginnote}

%Raijamakers. Currently dominated by priors.

In addition to radius measurements, there has long been the possibility~\cite{lyne04,lattimer05} of obtaining a neutron-star moment of inertia measurement based on long-term radio timing of PSR J0737-3039, a binary pulsar system in which the periastron advance receives a small correction from relativistic spin-orbit coupling. A recent analysis~\cite{hu20} has shown that by 2030 a moment of inertia measurement of PSR J0737-3039A to 11\% precision at the 68\% confidence level will be achievable. The moment of inertia for a uniformly rotating neutron star of radius $R$ and angular velocity $\Omega$ can be calculated in the slow-rotation approximation, valid for most millisecond pulsars, by solving the %normal 
TOV equations together with
\begin{equation}
I = \frac{8\pi}{3}\int_0^R r^4 (\varepsilon + p)e^{(\lambda -\nu)/2}\,
\frac{\bar\omega}{\Omega} \,\dd{r},\, \hspace{.2in} e^{-\lambda} = \left(1 - \frac{2m}{r}\right)^{-1}, \hspace{.2in} 
\dv{\nu}{r} = -\frac{2}{\varepsilon + p} \dv{p}{r}\,,
\end{equation}
where $\lambda$ and $\nu$ are metric functions and $\bar \omega$ is the rotational drag. In Refs.\ \cite{lim19prc,greif20} the moment of inertia of PSR J0737-3039A, which has a very well measured mass of $M=1.338\Msun$, was calculated from EOSs based on \ChEFT. In Ref.\ \cite{lim19prc} it was found that at the 95\% credibility level, the moment of inertia of J0737-3039A lies in the range $0.98 \times 10^{45}$\,g\,cm$^{2} < I < 1.48 \times 10^{45}$\,g\,cm$^{2}$, while Ref.\ \cite{greif20} found a consistent but somewhat larger range of $1.06 \times 10^{45}$\,g\,cm$^{2} < I < 1.70 \times 10^{45}$\,g\,cm$^{2}$. The moment of inertia is strongly correlated with the neutron star radius, and it has been shown \cite{raithel16} that measurements of the PSR J0737-3039A moment of inertia can constrain its radius to within $\pm 1$\,km.
%The crustal component of the moment of inertia is also strongly correlated with the amount of superfluid angular momentum in the inner crust, and in order to explain the largest glitches observed in the Vela pulsar, a large angular momentum reservoir is needed~\cite{andersson12,chamel13,piekarewicz14}. Modeling based on \ChEFT~\cite{lim19prc} suggests that the crustal component of the moment of inertia is less than 7\%, which would be too small to account for glitch observations if free neutrons in the inner crust are strongly entrained. Relaxing the strong neutron entrainment scenario, \eg, by including more consistently the effects of pairing~\cite{watanabe17}, may be necessary to resolve the current discrepancy between microscopic \ChEFT calculations and inferred constraints on crustal moments of inertia from large pulsar glitches.
%\begin{marginnote}
%The radius and moment of inertia of a typical $1.4\Msun$ neutron star is strongly correlated with the value of \mbox{$E_{\rm sym}(n\approx 2n_0)$~\cite{lattimer01}}.
%, especially its value around twice saturation density $2n_0$. 
%\end{marginnote}

In the past ten years, several neutron stars~\cite{demorest2010,antoniadis2013,cromartie19} with well measured masses of $M\gtrsim 2\Msun$ have been observed. The maximum mass ($M_{\rm max}^{\rm TOV}$) of a non-rotating neutron star is a key quantity to probe the composition of the inner core, which must have a sufficiently stiff EOS to support the enormous pressure due to the outer layers.\footnote{Beyond $M_{\rm max}^{\rm TOV}$, additional stable branches~\cite{alford13}, such as hybrid quark-hadron stars or pure quark stars, may appear before the ultimate collapse to a black hole.} To date the strongest candidate for the heaviest measured neutron star is PSR J0740+6620, with a mass of $M=2.14^{+0.20}_{-0.18}\Msun$ at the 95\% credibility level~\cite{cromartie19}. As mentioned previously, \ChEFT generically gives rise to relatively soft EOSs just above nuclear saturation density. The existence of a very massive neutron star with $M=2.14\Msun$ would require a stiff EOS at high densities, revealing a slight tension with \ChEFT~\cite{drischler20c}. However, even smooth extrapolations~\cite{lim18,Huth:2020ozf} of EOSs from \ChEFT can produce maximum neutron star masses in the range $2.0\Msun \lesssim M_{\rm max}^{\rm TOV} \lesssim 2.4\Msun$, and therefore more precise radius measurements (or the observation of heavier neutron stars) are needed to make strong inferences about the EOS in the \ChEFT validity region $n\lesssim 2n_0$.
%
%\begin{marginnote}
%Beyond $M_{\rm max}^{\rm TOV}$, additional stable branches~\cite{alford13}, such as hybrid quark-hadron stars or pure quark stars, may appear before the ultimate collapse to a black hole.
%\end{marginnote}

\subsection{Neutron star mergers}

The advent of gravitational wave astronomy has opened a new window into the visible Universe. Current gravitational wave detectors (LIGO and Virgo) are sensitive to frequencies $10\,{\rm Hz} < f < 10\,{\rm kHz}$, which is the prime range for compact object mergers and supernovae. Gravitational wave astronomy therefore has major implications for the field of nuclear astrophysics~\cite{baiotti19}. In particular, during the late-inspiral phase of binary neutron star coalescence, a pre-merger neutron star will deform with induced quadrupole moment $Q$ under the large tidal gravitational field ${\cal E}$: $Q_{ij} = - \lambda {\cal E}_{ij}$,
where $\lambda$ is the dimensionful tidal deformability parameter. Tidal deformations enhance gravitational radiation and increase the rate of inspiral. Gravitational wave detectors are sensitive to such phase differences and hence the dense matter EOS, but such corrections enter formally at fifth order~\cite{flanagan08} in a post-Newtonian expansion of the waveform phase and are therefore difficult to extract. The tidal deformability is an important observable in its own right, but this quantity is also strongly correlated with both the neutron star radius~\cite{annala18}, since more compact stars experience a smaller deformation under a given tidal field, and especially the moment of inertia through the celebrated I-Love-Q relations~\cite{yagi13}. The post-merger gravitational wave signal from binary neutron star coalescence can also carry important information on the nuclear EOS. It has been shown~\cite{bauswein12} that the peak oscillation frequency $f_\mathrm{peak}$ of a neutron-star merger remnant is strongly correlated with neutron star radii. Moreover, a strong first-order phase transition can show up as a deviation in the empirical correlation band between $f_\mathrm{peak}$ and $\Lambda$~\cite{bauswein19}.
\begin{marginnote}
	\entry{LIGO}{Laser Interferometer Gravitational-Wave Observatory}
\end{marginnote}

\begin{figure}[tb]
        \begin{center} \includegraphics[width=1.0\textwidth]{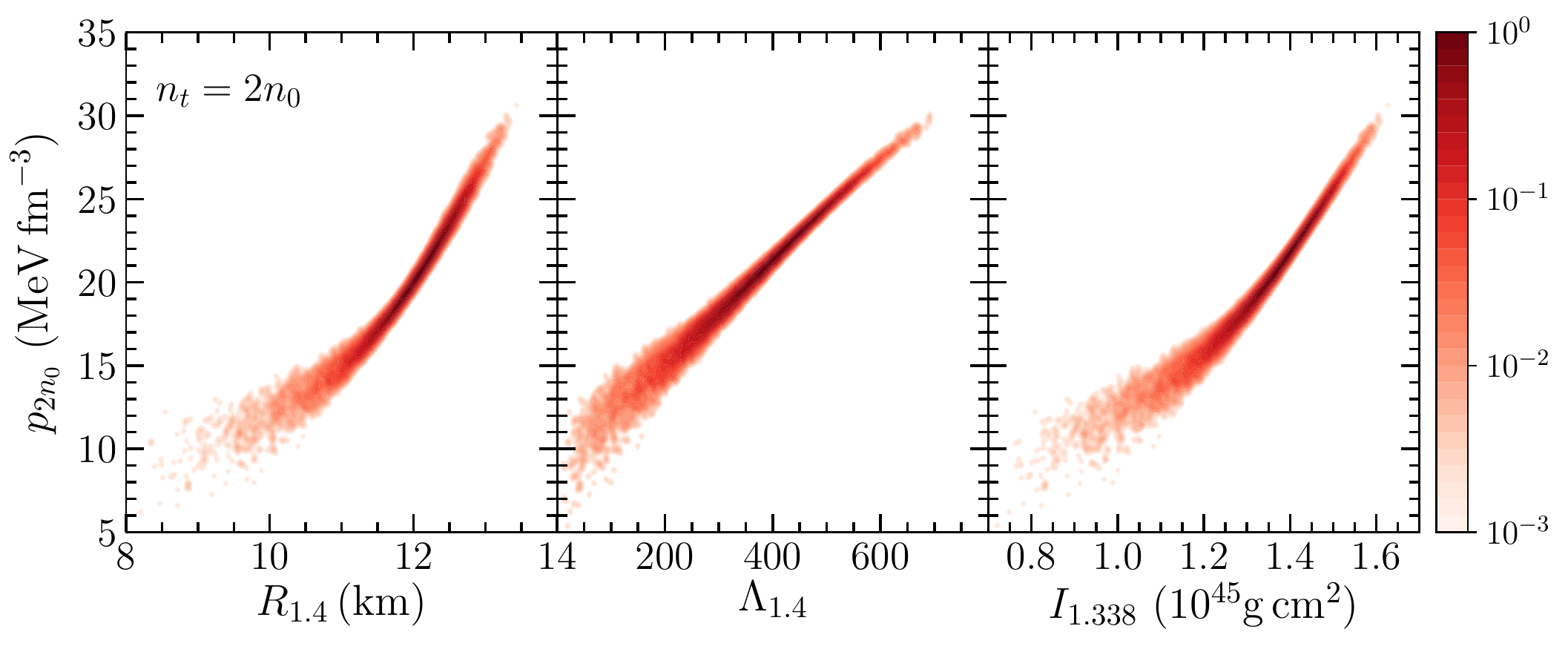}
        \end{center}
        \caption{
        Joint probability distribution for the pressure of beta-equilibrium matter 
        %at twice saturation density 
        $p(2n_0)$ with the radius of a $1.4\Msun$ neutron star (left), the tidal deformability of a $1.4\Msun$ neutron star (middle), and the moment of inertia of PSR J0737-3039A with a mass of $1.338\Msun$ (right). Results obtained from the Bayesian modeling of the nuclear EOS in Ref.\ \cite{lim18}.
        }
        \label{fig:press_ns}
\end{figure}

The first observation~\cite{abbott17a}, GW170817, of a neutron star merger through its gravitational wave emissions was accompanied by a short gamma-ray burst and optical counterpart~\cite{abbott17c}. The combined multi-messenger observations of this single event have resulted in a wealth of new insights about the origin of the elements and the properties of neutron stars. Analysis of the gravitational waveform resulted in a prediction \cite{abbott18} $\Lambda_{1.4} = 190^{+390}_{-120}$ for the dimensionless tidal deformability $\Lambda = \lambda / M^5$ of a $1.4\Msun$ neutron star. Theoretical predictions from \ChEFT~\cite{lim18,tews18lam} predating the analysis in Ref.\ \cite{abbott18} yielded similarly small tidal deformabilities $140<\Lambda<520$~\cite{lim18}. Analogous constraints on the binary tidal deformability parameter 
\begin{equation}
\tilde \Lambda = \frac{16}{13}\frac{(m_1+12 m_2)m_1^4 \Lambda_1 
                                                    + (m_2+12 m_1)m_2^4 \Lambda_2}{(m_1+m_2)^5}
\end{equation}
from gravitational-wave data [$\tilde \Lambda = 300^{+420}_{-230}$ \cite{abbott19}] and \ChEFT [$80 < \tilde \Lambda < 580$ \cite{tews18lam}] were similarly consistent. From the strong correlation between neutron star radii and tidal deformabilities, the LIGO/Virgo Scientific Collaboration reported~\cite{abbott18} an inferred constraint of $R=11.9^{+1.4}_{-1.4}\km$ for both of the neutron stars involved in the merger under the assumption that the shared EOS could support $2\Msun$ neutron stars. Only the combined mass $M_{\rm tot} = 2.74^{+0.04}_{-0.01}\Msun$ of the binary was very well measured from the gravitational waveform, but neither of the individual component masses $1.17\Msun < M_{1,2} < 1.60\Msun$ were expected~\cite{abbott17a} to deviate more than 20\% from the canonical value of $M\simeq 1.4\Msun$, assuming low neutron star spins. In summary, GW170817 data were found to strongly favor the soft EOSs predicted from \ChEFT, though many other models~\cite{fattoyev18} with generically stiffer EOSs were consistent with the upper bounds on $\Lambda$ and $R_{1.4}$ from GW170817.

Current gravitational wave interferometers do not have large signal-to-noise ratios at the high frequencies expected during the post-merger ringdown phase and therefore GW170817 provided no clues about the fate of the merger remnant. Nevertheless, an analysis of the spectral and temporal properties of the kilonova~\cite{metzger10} optical counterpart to GW170817 have been used~\cite{bauswein17,margalit17,shibata17,radice18,rezzolla18} to infer the lifetime of the merger remnant. Depending on the component neutron star masses prior to merger (primarily the total mass $M_{\rm tot}$) as well as the maximum mass for a nonrotating neutron star $M_{\rm max}^{\rm TOV}$, the merger remnant can (i)~undergo immediate collapse to a black hole, (ii)~exist as a short-lived hypermassive neutron star supported against collapse by differential rotation, (iii)~persist as a longer-lived supramassive neutron star supported against collapse by rigid-body rotation, or (iv)~form a stable massive neutron star. While there is still some uncertainty about what ranges of $M_{\rm tot}$ will lead to each of the above four scenarios, it has been suggested~\cite{margalit17} that prompt collapse will occur when $M_{\rm tot} \gtrsim 1.3-1.6 \,M_{\rm max}^{\rm TOV}$,  hypermassive neutron stars will be created when $1.2\,M_{\rm max}^{\rm TOV} \lesssim M_{\rm tot} \lesssim 1.3-1.6 \, M_{\rm max}^{\rm TOV}$, and supramassive neutron stars will result when $M_{\rm tot} \lesssim 1.2 \, M_{\rm max}^{\rm TOV}$. Each merger outcome is expected to have a qualitatively different optical counterpart and total mass ejection, since longer remnant lifetimes generically give rise to more and faster moving disk wind ejecta.

Observations of the GW170817 kilonova suggest that the most likely outcome of the neutron star merger was the formation of a hypermassive neutron star, which would imply a value of $M_{\rm max}^{\rm TOV} = 2.15-2.35\Msun$~\cite{shibata17,margalit17,rezzolla18}. Eliminating the possibility of prompt black hole formation in GW170817 also rules out compact neutron stars with small radii and tidal deformabilities. In Ref.~\cite{bauswein17} such arguments were used to infer that the radius of a $1.6\Msun$ neutron star must be larger than $R_{1.6} \gtrsim 10.7\km$, while in Ref.~\cite{radice18} it was found that the binary tidal deformability parameter for the GW170817 event must satisfy $\tilde \Lambda \gtrsim 400$. Both of these inferred constraints are compatible with predictions~\cite{lim18,tews18lam,lim19epj,tews19epj} from \ChEFT. However, the constraint on the binary tidal deformability $\tilde \Lambda > 400$ can rule out a significant set of soft EOSs~\cite{capano19}, roughly half of those allowed in the analysis of Ref.~\cite{lim18}. Combined gravitational wave and electromagnetic observations of binary neutron star mergers together with more precise radius measurements therefore have the possibility to strongly constrain the dense matter EOS and related neutron star properties in the regime of validity of \ChEFT~\cite{capano19,lim20,Raaijmakers:2019dks}. As a demonstration, in Fig.~\ref{fig:press_ns} we show the joint probability distributions~\cite{lim18} for the pressure of nuclear matter at $n=2n_0$ with (i)~the radius of a $1.4\Msun$ neutron star, (ii)~the tidal deformability of a $1.4\Msun$ neutron star, and (iii)~the moment of inertia of PSR J0737-3039A with a mass of $1.338\Msun$. 

%Space for GW190814 and $\Gamma_{\rm th}$?

\subsection{Core-collapse supernovae}
Neutron stars are born following the gravitational collapse and ensuing supernova explosion of massive stars ($M \gtrsim 8 \Msun$). The core bounce probes densities only slightly above normal nuclear saturation density~\cite{oertel17} and leaves behind a hot ($T\sim 20-50 \MeV$) nascent proto-neutron star. During the subsequent Kelvin-Helmholtz phase that lasts tens of seconds, the proto-neutron star emits neutrinos, cools to temperatures $T < 5 \MeV$, de-leptonizes, and contracts to reach supra-saturation densities in the innermost core. The success or failure of the supernova itself~\cite{janka07}, the thermal and chemical evolution during the Kelvin-Helmholtz phase~\cite{pons99}, and the possibility of novel nucleosynthesis in the neutrino-driven wind~\cite{roberts10} depend on details of the nuclear EOS and weak reaction rates. 
%In this section we will focus on the prominent role of nuclear interactions on various theoretical inputs to hydrodynamical simulations of core-collapse supernovae.

Investigating the qualitative impact of specific EOS properties, 
%empirical parameters,
such as the incompressibility or the symmetry energy, on the fate of supernova explosions is often challenging due to ``Mazurek's Law'', a colloquial observation that feedback effects tend to wash out any fine tuning of parameters in core-collapse supernovae~\cite{mueller16}. Nevertheless, several recent systematic investigations~\cite{schneider19,yasin20} of EOS parameters have supported the idea that a high density of states, linked to a large value of the in-medium nucleon effective mass $M^*$, reduces thermal pressure and leads to enhanced contraction of the initial proto-neutron star. This results in the emission of higher-energy neutrinos that support the explosion through the neutrino reheating mechanism~\cite{janka07}. Since microscopic calculations based on \ChEFT tend to predict larger values of the effective mass than many mean-field models~\cite{rrapaj16,Carbone:2019pkr}, these observations help motivate recent efforts~\cite{du19,Huth:2020ozf} to include thermal constraints from \ChEFT directly into supernova EOS tables. 
%(Could add some discussion about $\Gamma_{\rm th}$ here.)

%Modern simulations of core-collapse supernovae have identified neutrino reheating as an essential component of successful supernova explosions \cite{janka07}, since the energy imparted to the initial shockwave through core bounce is typically insufficient to result in a prompt supernova explosion.
Neutrino reactions also affect the nucleosynthesis outcome in neutrino-driven wind outflows and the late-time neutrino signal that will be measured with unprecedented detail during the next galactic supernova. Charged-current neutrino-absorption reactions $\nu_e+n \rightarrow p + e^-$ and $\bar \nu_e+p \rightarrow n + e^+$, which can be calculated from the imaginary part of vector and axial vector response functions, are especially sensitive~\cite{roberts12,martinez12} to nuclear interactions and in particular the difference $\Delta U = U_n - U_p$ between proton and neutron mean fields. The isovector mean field is especially important in the neutrinosphere, the region of warm and dense matter where neutrinos decouple from the exploding star. Recently, the calculation~\cite{rrapaj15,rrapaj16} of nuclear response functions that include mean-field effects from \ChEFT interactions have shown that terms beyond the Hartree-Fock approximation are needed for accurate modeling. In particular, resummed particle-particle ladder diagrams were shown~\cite{rrapaj15} to produce larger isovector mean fields due to resonant, non-perturbative effects in the NN interaction. Moreover, for neutral-current neutrino reactions, such as neutrino pair bremsstrahlung and absorption, resonant NN interactions were shown to significantly enhance reaction rates at low densities compared to the traditional one-pion exchange approximation~\cite{bartl14}.

%\begin{afterthoughts}[Afterthoughts: remove prior to submission]
%	\begin{enumerate}
%		\item Does it make sense to move the QCD phase diagram to the Introduction?
%		\item The (very nice!) summary plot of the symmetry energy constraint would fit well in Section~4. The same Bayesian results will be shown in another plot as well. Would that change a lot?
%		\item Discuss what constraints pulsar masses, GWs, and NICER results actually put on the EOS. Density regime. Discuss how far the EOS predicted from chiral EFT agrees with observational and experimental constraints---and vice versa. What are the most stringent results? Where to go next? What to expect?
%		\item The title of the subsection ``Neutron stars'' seems to be too broad.
%	\end{enumerate}
%\end{afterthoughts}

%\begin{summary}[SUMMARY POINTS]
%	\begin{enumerate}
%		\item \ChEFT can place meaningful constraints on the cold dense matter EOS.
%		\item \ChEFT provides a controlled theoretical framework for constraining the nature of hot and dense matter in supernovae, neutron stars, and neutron star mergers. \com{needs work}
%	\end{enumerate}
%\end{summary}

%% file: sections/05_summary_outlook.tex
\section{Summary and outlook} \label{sec:summary_outlook}

In this article, we have reviewed recent progress in \ChEFT calculations of nuclear matter properties
% at densities $n \lesssim 2 n_0$ 
(with quantified uncertainties) and their
% applications to neutron stars through high-density extrapolations. 
implications in the field of nuclear astrophysics. %through high-density extrapolations. 
Combined with observational and experimental constraints, these microscopic calculations provide the basis for improved modeling of 
%the construction of the high-density EOS relevant for
%the hot and dense matter present in 
supernovae, neutron stars, and neutron star mergers.
% through high-density extrapolations. 
In particular, we have highlighted MBPT as an efficient framework for studying the nuclear EOS and transport properties %
%at varying {\color[rgb]{0,0,1}(across a wide range of?)} 
across a wide range of densities, isospin asymmetries, and temperatures. %zero and finite temperature with chiral interactions up to \NNNLO. %, which can be softened using RG methods to improve the many-body convergence. 
We have also shown how advances in high-performance computing have enabled the  implementation of two- and multi-nucleon forces in MBPT up to high orders in the chiral and many-body expansions. 
%A fully automated Monte Carlo approach to MBPT has recently become available due to the advances made in the implementation of chiral many-body forces, the efficient derivation and evaluation of MBPT diagrams, and high-performance computing. 
%This has led to MBPT calculations at high orders in the chiral and many-body expansion without needing the common approximations. 
Finally, we have described new tools for
%the importance of 
quantifying theoretical uncertainties (especially EFT truncation errors) to confront 
microscopic calculations of the nuclear EOS 
%the nuclear EOS %obtained from different many-body frameworks 
%and nuclear interactions 
with empirical constraints. Such systematic studies
%between different theory predictions and empirical constraints 
are particularly important in view of  %observational and experimental 
EOS constraints %on the neutron star EOS 
anticipated in the new era of multimessenger astronomy;
%(\eg, tidal deformabilities from the LIGO-Virgo collaboration and simultaneous mass-radius measurements of neutron stars from NICER) 
\eg, from gravitational wave detection, mass-radius measurements of neutron stars, and experiments with neutron-rich nuclei.
In the following we briefly summarize %our perspectives on future nuclear matter calculations and their applications to nuclear astrophysics.
several open research directions at the interface of nuclear EFT and high-energy nuclear astrophysics.
(i)~EFT truncation errors and their correlations in density and across observables need to be studied with different many-body frameworks and nuclear interactions at arbitrary isospin asymmetry and finite temperature.
%The extension to arbitrary isospin asymmetries and finite temperatures is also important.
%
(ii)~Together with EFT truncation errors, the uncertainties in the LECs parametrizing the interactions need to be quantified and propagated to nuclear matter properties using a comprehensive Bayesian statistical analysis.
%of the nuclear EOS can be performed via Monte Carlo sampling over GP hyperparameters and the LECs in the nuclear interactions.
%
(iii)~The full uncertainty quantification of the nuclear EOS will be aided by the development of improved order-by-order chiral NN and 3N potentials and the study of different regularization schemes as well as delta-full chiral interactions. 
%
%(iii)~Constructing a microscopically constrained EOS for astro simulations/applications
%
%(iv)~or applications to astrophysical simulations of neutron-stars and supernovae,  an important subject of future research consists in extending  theoretical calculations to consistently include non-nucleonic particles at finite $T$, such as pions and hyperons. %%mention also hyperon puzzle?
%
(iv)~Many-body calculations of nuclear matter properties beyond the nuclear EOS (\eg, linear response and transport coefficients) with chiral NN and 3N interactions are required for more accurate numerical simulations of supernovae and neutron star mergers.
(v)~Future neutron star observations will provide stringent %constraints 
tests of nuclear forces and nuclear many-body methods in a regime that is presently largely unconstrained.
%the EOS and composition of matter at high density as well as the nature of nuclear forces in a regime that is presently largely unconstrained. additional insights into the important open question up to which density \ChEFT is useful to constrain the nuclear EOS. 
%, and may possibly also allow one to probe  \ChEFT's efficacy at densities $n \gg n_0$. Future observations can be expected to provide stringent constraints on the high-density EOS, and may possibly also allow one to probe  \ChEFT's efficacy at densities $n \gg n_0$. 
The interplay between observation, experiment, and theory in the next decade can be expected to result in many further advances in our understanding of strongly interacting matter.